\documentclass[journal,twoside,web]{ieeecolor}
\usepackage{generic}
\usepackage{graphicx}
\usepackage{color}          %
\usepackage{amsfonts}       %

\usepackage{amsthm}         %
\usepackage{mathtools} 
\graphicspath{{figures/}}

\definecolor{Granata}{rgb}{0.64,0,0} 

\newcommand{\mc}[1]{\mathcal{#1}}
\newcommand{\mb}[1]{\mathbb{#1}}
\newcommand{\N}{N}

\newcommand{\poa}{{\rm PoA}}
\newcommand{\nashe}[1]{{\rm NE}({#1})}
\newcommand{\be}{\begin{equation}}
\newcommand{\ee}{\end{equation}}
\newcommand{\geefwn}{\mathcal{G}_{f,w}^n}

\renewcommand{\ae}{a^{\rm ne}}

\newcommand{\aopt}{a^{\rm opt}}
\newcommand{\ami}{a_{-i}}
\newtheorem{definition}{Definition}
\DeclareSymbolFont{bbold}{U}{bbold}{m}{n}
\DeclareSymbolFontAlphabet{\mathbbold}{bbold}
\newcommand{\vect}[1]{\mathbbold{#1}}
\newcommand{\ones}[1][]{\vect{1}_{#1}}

\def\eqspacezero{\hspace*{-0.3mm}}

\def\eqspacetwo{\hspace*{-0.08mm}}

\newcommand{\fopt}{f^{\rm opt}}
\newcommand{\muopt}{\mu^{\rm opt}}

\let\labelindent\relax
\usepackage{enumitem}
\usepackage{cite}
\usepackage{tikz}
\usepackage{pgfplots}
\usetikzlibrary{intersections,backgrounds,plotmarks}

\newtheorem{theorem}{Theorem}
\newtheorem{assumption}{Assumption}
\newtheorem{remark}{Remark}
\newtheorem{corollary}{Corollary}

\newtheorem{proposition}{Proposition}

\def\myspace{1.5mm}
\def\myspaceintro{1.5mm}
\newcommand{\cdotshort}{\!\cdot\!}
\newcommand{\fsv}{f^{\rm es}}
\newcommand{\fmc}{f^{\rm mc}}
\newcommand{\Wsv}{W^{\rm es}}
\newcommand{\Wmc}{W^{\rm mc}}

\newtheoremstyle{break}
  {\topsep}{\topsep}%
  {\itshape}{}%
  {\bfseries}{}%
  {\newline}{}%
\theoremstyle{break}

\usepackage{amsmath}

\DeclareMathOperator*{\argmax}{arg\,max}
\DeclareMathOperator*{\argmin}{arg\,min}

\definecolor{PortlandGreen}{RGB}{99,166,63}
\definecolor{MyGreen}{RGB}{0,180,0}
\colorlet{green}{MyGreen}

\newcommand{\cdue}{C_1}
\newcommand{\ctre}{C_2}
\newcommand{\cquattro}{C_3}
\renewcommand{\k}{k}

\usepackage{ctable} 
\usepackage{algorithm}
\usepackage{algpseudocode}
\algnewcommand{\LeftComment}[1]{\Statex \(\triangleright\) #1} %
\newcommand{\coloneqq}{=}

\usepackage{xr}
\begin{document}
\colorlet{myred}{black}

\title{\LARGE \bf Utility Design for Distributed Resource Allocation -- Part II:\\
\mbox{Applications to Submodular, Covering, and Supermodular Problems}
}
\author{Dario~Paccagnan~and~Jason~R.~Marden
\thanks{
This research was supported by SNSF Grant \#P2EZP2-181618 and by ONR Grant \#N00014-15-1-2762, NSF Grant \#ECCS-1351866. D. Paccagnan is with the Department of Computing, Imperial College London, London SW72AZ, U.K. (e-mail: {\tt d.paccagnan@imperial.ac.uk}). J.\,R. Marden is with the Department of Electrical and Computer Engineering, and the Center of Control, Dynamical Systems and Computation, UC Santa Barbara, CA 93106-5070, USA (email: {\tt jrmarden@ucsb.edu}).}
}
\maketitle

\begin{abstract}
A fundamental component of the game theoretic approach to distributed control is the design of local utility functions.
Relative to resource allocation problems that are additive over the resources, Part I showed how to design local utilities so as to maximize the associated performance guarantees \cite{part1Paccagnan2018}, which we measure by the price of anarchy. 
The purpose of the present manuscript is to specialize these results to the case of submodular, covering, and supermodular problems.
In all these cases we obtain tight expressions for the price of anarchy that often match or improve the guarantees associated to state-of-the-art approximation algorithms.
Two applications and corresponding numerics are presented: the vehicle-target assignment problem and a coverage \mbox{problem arising in wireless data caching.}
\\[3mm]

\end{abstract}

\section{Introduction}
\IEEEPARstart{T}{here} has been a growing interest in recent years in the analysis and control of multi agent and networked systems. The potential of such systems stems from the societal impact that they promise to deliver: from medicine \cite{ishiyama2002magnetic} to surveillance \cite{srivastava2013stochastic}, from  future mobility \cite{spieser2014toward} to food production \cite{kozai2015plant}, to name a few. The typical challenge in the control of such systems is the design of agent-level decision rules that are capable of achieving a desirable joint objective by relying solely on local information. A recently developed approach based on game theoretic arguments and termed \emph{game design} \cite{shamma2007cooperative} has proved useful in complementing the results obtained by more traditional techniques. This approach amounts to assigning a local utility function to each agent so that their self interested maximization recovers the desired system level objective, i.e., jointly maximizes a given system-level~objective. 
   
In Part I of this work we applied this approach to a class of combinatorial resource allocation problems, where a finite number of agents need to be allocated to a set of resources, with the goal of maximizing a given welfare function, additive over the resources. In this context, the notion of price of anarchy (the ratio between the welfare at the worst-performing Nash equilibrium and the optimum welfare) was used as the performance metric. Indeed, any algorithm capable of computing a Nash equilibrium inherits an approximation ratio equal to the price of anarchy. %
More specifically, we tackled the utility design problem:
\begin{itemize}
\item[i)]  Given local utility functions, how do we characterize the price of anarchy?
\item[ii)] Is it possible to select utility functions so as to maximize such performance metric?	
\end{itemize}

\noindent While Part I showed how to compute and optimize the price of anarchy ($\poa$) by means of tractable linear programs, this manuscript specializes these results to the case of submodular, supermodular and covering problems. 

\vspace*{\myspaceintro}
\noindent{\bf Contributions.} The following highlights our contributions: 
\setlist[enumerate]{leftmargin=*}
\begin{enumerate}
\item Relative to monotone submodular problems, we provide the analytical expression of the $\poa$ as a function of the utilities assigned to every agent (Theorem \ref{thm:wconcavefwdecreas}). 
To the best of our knowledge, this is the first result that gives an exact characterization of the $\poa$ for this class of problems.
We specialize this result to the Shapley value and marginal contribution mechanisms, recovering a partial result presented in \cite{marden2014generalized}.
Finally, we show how the performance certificates offered by this approach improve on state-of-the-art approximation algorithms.
\item Relative to covering problems, we derive the exact expression for the $\poa$ as the maximum between $\mc{O}(n)$ numbers, where $n$ is the number of agents in the system (Theorem \ref{thm:poageneralcovering}). This result strengthens previous findings in \cite{gairing2009covering,paccagnan2017arxiv}, which required additional assumptions on the structure of the utility functions. Optimal utilities yield a $\poa$ of $1-1/e$, where $e$ is Euler's number.
\item Relative to supermodular problems, we provide the expression for the $\poa$ (Theorem \ref{thm:convex}), complementing previous bounds appearing in \cite{phillips2017design, jensen2018}.
\item We present two applications demonstrating how our approach yields improved numerical performances.
\end{enumerate}

\vspace*{\myspaceintro} 
\noindent {\bf Organization.} %
Section \ref{sec:problemformulation} introduces the problem formulation, the game theoretic approach, and the two main results contained in Part I. In Sections \ref{sec:submod}, \ref{sec:coveringg} and \ref{sec:supermod} 
we specialize the results of Part I to submodular, covering, supermodular problems, and present two corresponding applications. 
A discussion on the complexity of computing a pure Nash equilibrium is included.
\mbox{All the proofs are reported in Appendix \ref{appendixb}.}

\vspace*{\myspaceintro} 
\noindent {\bf Notation.} 
For any two positive integers $p\le q$, denote $[p]=\{1,\dots,p\}$ and $[p, q]=\{p,\dots,q\}$. We use $\mb{N}$, $\mb{R}_{>0}$ and $\mb{R}_{\ge0}$ to denote the set of natural numbers, positive and non-negative real numbers, respectively. We denote with $\ones[n]$ the vector with $n$ unit entries, and with $|\mc{I}|$ the cardinality of the finite set $\mc{I}$.
\section{Problem formulation and performance metrics}
\label{sec:problemformulation}
\subsection{Problem formulation}
Let $N=\{1,\dots,\k\}$ be a set of agents, and $\mc{R}=\{r_1,\dots,r_m\}$ be a set of resources, where each resource $r\in\mc{R}$ is associated with a value $v_r\ge 0$ describing its importance, $\k,m\in\mb{N}$. Every agent $i\in N$ selects a subset of the resources $a_i$ from a given collection $\mc{A}_i\subseteq 2^{\mc{R}}$, i.e., $a_i\in\mc{A}_i$. The welfare of an allocation $a = (a_1,\dots ,a_{\k})\in \mc{A} = \mc{A}_1 \times \dots \times \mc{A}_{\k}$ is 
\be
W(a)\coloneqq\sum_{r\in\cup_{i\in N} a_i}v_r w(|a|_r),
\label{eq:welfaredef}
\ee
where $w:[\k]\rightarrow \mb{R}$ is called the welfare basis function,  and $|a|_r\!=\!|\{i\in\N\,\text{s.t.}\,r\!\in\!a_i\}|$ denotes the number of agents selecting resource $r$ in allocation $a$. The system designer is interested in finding an optimal allocation, i.e.,
$\aopt\in\argmax_{a\in\mc{A}}W(a).
$
We denote the decisions of all agents but $i$ with $\ami=(a_1,\dots,a_{i-1},a_{i+1}, \dots, a_{\k})$.
\subsection{A game theoretic approach}
We follow the game theoretic approach in \cite{part1Paccagnan2018} to obtain a distributed solution to the previous problem, and assign to each agent $i\in\N$ a local utility function $U_i:\mc{A}\rightarrow\mb{R}$ of the form 
\be
U_i(a) = \sum_{r\in a_i}v_r f(|a|_r)\,,
\label{eq:utilities}
\ee
where $f:[\k]\rightarrow \mb{R}$ is our \emph{design choice}; we refer to it as to the \emph{utility generating mechanism}, or simply the mechanism, since each agent's utility is fully determined once $f$ is specified.
We identify the game introduced above with the tuple 
\begin{equation}	
G=(N, \mc{R}, \mc{A}, \{v_r\}_{r\in \mc{R}}, w, f)\,,
\label{eq:gameG}
\end{equation}
and introduce the following family of games.
\begin{definition} 
\label{def:ass}
Given $n\in\mb{N}$, a welfare basis $w:[n]\rightarrow\mb{R}_{>0}$ and a mechanism $f:[n]\rightarrow\mb{R}$, we let $\geefwn$ be the set containing all games $G$ of the form \eqref{eq:gameG}, where $\mc{R}$ is any set of resources, $\mc{A}$ is any allocation set, $\{v_r\}_{r\in\mc{R}}$ is any tuple of non-negative resource values, and
\begin{itemize}
	\item[i)] the number of agents is upper bounded by $|N|\le n$,
	\item[ii)] the optimum value satisfies $W(\aopt)>0$. 
\end{itemize}
\end{definition}
\noindent Given a mechanism $f$, we measure its performance on the class $\geefwn$ adapting the notion of \emph{price of anarchy} \cite{Koutsoupias} as
\be
\poa(f,w,n) = \inf_{G\in \geefwn}\biggl(\frac{\min_{a\in \nashe{G}} W(a)}{\max_{a\in\mc{A}} W(a)}\biggr)\,,
\label{eq:poadef}
\ee
where $\nashe{G}$ denotes the set of pure Nash equilibria of $G$. By definition $0\le \poa(f,w,n)\le 1$, and the higher the price of anarchy, the better performance certificates we can offer. Observe how the requirement $W(\aopt)>0$ is solely needed to ensure that $\poa(f,w,n)$ is well defined. Additionally note that any game with utility functions given in \eqref{eq:utilities} is a congestion game (see Appendix \ref{app:congestion}), therefore ensuring existence of a pure Nash equilibrium \cite{monderer1996potential}.
Since $w$ and $f$ associate a real number to every integer number in $\{1,\dots,n\}$, we often denote $f$ and $w$ as vectors in $\mb{R}^n$.

It is important to observe that any equilibrium allocation $a^{\rm ne}\in\nashe{G}$ yields a welfare that is at least a fraction $\poa(f,w,n)$ of $W(\aopt)$ over all instances $G\in\geefwn$,~i.e., 
\[
W(a^{\rm ne})\ge {W(\aopt)} \cdot \poa(f,w,n)\,.
\]
In other words, the price of anarchy represents the \emph{approximation ratio} of any algorithm capable of computing a pure Nash equilibrium. For this reason, the price of anarchy will serve as the performance metric in all the forthcoming analysis. 

For sake of completeness, we conclude this section including one of the most commonly used equilibrium-computing algorithms: the round robin \emph{best response dynamics}.
\newlength\algbrake
\setlength\algbrake{2mm}

\begin{algorithm}[h!]
\caption{Best response dynamics (round-robin)}\label{alg:BR}
\begin{algorithmic}[1]
\State Initialise $a^0\in\mc{A}$;\quad $t\gets 0$
\vspace*{\algbrake}
\While {not converged}
\LeftComment{Best response}
\State $i\gets (t \mod n)+1$
\State $a^{t+1}_i\gets \argmax_{a_i\in\mc{A}_i} U_i(a_i,\ami^t)$ 
\State  $a^{t+1}\gets (a^{t+1}_i,\ami^t)$
\State $t\gets t+1$
\EndWhile
\end{algorithmic}
\end{algorithm}
\noindent 
With a single round of the best response dynamics we identify the process where all players update their decision once, in a given order, and ties are broken according to a pre-specified rule. We remark that Algorithm \ref{alg:BR} is \emph{guaranteed} to converge to a pure Nash equilibrium in a finite number of rounds for \emph{any} possible instance $G$ as $G$ is a congestion game \cite{monderer1996potential}. Similar convergence results hold almost surely if the players updating their decision are, e.g., uniformly randomly selected from $[n]$. This will produce a totally asynchronous algorithm.

\subsection{Two results from Part I}
\label{subsec:tworesults}
In Part I, we showed how to compute and optimize $\poa(f,w,n)$ through the solution of tractable linear programs \cite{part1Paccagnan2018}. Two results presented in Part I and needed in the development of Part II are summarized in the following theorem. Before doing so, we introduce a useful set of integer tuples
\[\begin{split}
\mc{I}_R\coloneqq\{&(a,x,b)\in[0,n]^3~\text{with}~1
\le a+x+b\le n\\
&\text{s.t.}~ a\cdot x\cdot b=0~\text{or}~a+x+b=n\}\,.
\end{split}
\]
Informally, $\mc{I}_R$ contains all the integer tuples $(a,x,b)$ lying on the surface of a tetrahedron with vertices $(0,0,0)$, $(n,0,0)$, $(0,n,0)$, $(0,0,n)$, excluding the tuple $(0,0,0)$. %
\begin{figure}[b!]
\begin{center}
\vspace*{-8mm}
\includegraphics[scale=0.65]{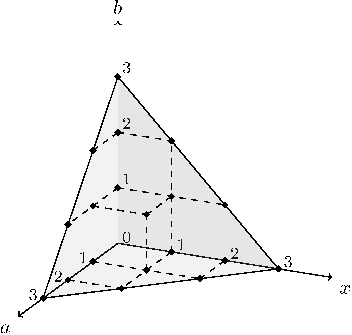}
\caption{The diamonds represent the points belonging to $\mc{I}_R$,~$n=3$.}
\vspace*{-3mm}
\label{fig:checkered}
\end{center}
\end{figure}

\noindent 
\newpage
\begin{theorem}[{\bf Characterizing and optimizing }$ \rm{\bf PoA}$, \cite{part1Paccagnan2018}]
\label{thm:mine}
\begin{itemize}
\item[]
\item[i)]
Let $w\in\mb{R}^n_{>0}$ be a welfare basis function, and let $f\in\mb{R}^n$.
$\bullet$ If $f(1)\le0$, then $\poa(f,w,n)=0$ for any $n\in\mb{N}$.\\
$\bullet$ If instead $f(1)>0$, $n\in \mb{N}$, the price of anarchy \eqref{eq:poadef} is given by 
$
\poa(f,w,n) = 1/{W^\star},
$ with $W^\star$ the value of the following (dual) linear program
	\be
	\label{eq:proppart1}
	\small
	\begin{split}
		&W^\star = \min_{\lambda\in\mb{R}_{\ge0},\,\mu\in\mb{R}}~ \mu \\[0.1cm]
		&\,\text{s.t.} ~w(b\eqspacezero  +\eqspacezero  x)
		\eqspacezero-\eqspacezero \mu w(a  \eqspacezero+\eqspacezero  x)\eqspacezero+  \eqspacezero\lambda[af(a\eqspacezero + \eqspacezero x)-bf(a  \eqspacezero+ \eqspacezero x\eqspacezero  + \eqspacezero 1)]
		\!\le\! 0\\[0.1cm]
		& \hspace*{60mm}\forall (a,x,b)\in\mc{I}_R,
	\end{split}
	\ee
	where $f(0)=w(0)=f(n+1)=w(n+1)=0$.
	\vspace*{\myspace}
	
\item[ii)]
Let $w\in\mb{R}^n_{>0}$ be a welfare basis function, $n\in\mb{N}$.
The design problem %
	$
	\argmax_{f\in\mb{R}^n} \poa(f,w,n)
	$
	is equivalent to the following linear program %
	in $n+1$ scalar unknowns%
\be
\small
\label{eq:proppart2}
	\begin{split}
	&(\fopt,\muopt) \in \argmin_
{
\substack{f\in\mb{R}^n\\ f(1)\ge 1}
,\,\mu\in\mb{R}}
	~ \mu \\[0.1cm]
	&\,\text{s.t.}\,w(b \eqspacetwo   +  \eqspacetwo   x)
	 \eqspacetwo- \eqspacetwo \mu w(a  \eqspacetwo   + \eqspacetwo    x)  \eqspacetwo+af(a \eqspacetwo    +    \eqspacetwo x) \eqspacetwo- \eqspacetwo bf(a    +    \eqspacetwo x    \eqspacetwo +  \eqspacetwo   1) \eqspacetwo
	\le  \eqspacetwo0\\[0.1cm]
& \hspace*{60mm}\forall (a,x,b)\in\mathcal{I}_R,
	\end{split}
\ee
where $f(0)=w(0)=f(n+1)=w(n+1)=0$.
\end{itemize}
\end{theorem}

 In the forthcoming sections we specialize these general statements and obtain analytical results for the case of submodular, supermodular and maximum coverage problems. 
Towards this goal, we will often assume that $w:[n]\rightarrow\mb{R}_{>0}$ and $f:[n]\rightarrow\mb{R}$ satisfy the following assumption.
\begin{assumption}
\label{ass:fw=1}
	The welfare basis $w:[n]\rightarrow\mb{R}_{>0}$ and the {mechanism $f:[n]\rightarrow\mb{R}$ are normalized, i.e., $w(1)\!=\!f(1)\!=\!1$.}
\end{assumption}
\noindent We remark that Assumption \ref{ass:fw=1} is without loss of generality. Indeed,  for any given welfare basis $w:[n]\rightarrow\mb{R}_{>0}$, not necessarily satisfying $w(1)=1$, one can consider its normalized version $\bar w(j)=w(j)/w(1)$, $j\in[n]$,  which satisfies Assumption \ref{ass:fw=1} and yields the same price of anarchy of the original $w$.\footnote{Rescaling $w$ has the sole effect of rescaling the welfare $W(a)$ for all allocations $a\in\mc{A}$ with the same positive constant $w(1)$, thus leaving any optimal allocation $\aopt$ and any pure Nash equilibrium $\ae$ unchanged.} 
Aside for the case when $f(1)\le 0$ (for which $\poa(f,w,n)=0$ by Theorem \ref{thm:mine}), the same argument can be made for any $f:[n]\rightarrow\mb{R}$. It is indeed straightforward to verify that for any $f:[n]\rightarrow\mb{R}$ with $f(1)>0$, the normalized mechanism $\bar f(j)=f(j)/f(1)$, $j\in[n]$, satisfies Assumption \ref{ass:fw=1} and gives the same price of anarchy of $f$ (see \cite[Lem. \ref{lem:rescalingdoesnotchange}]{part1Paccagnan2018}).

\subsection{Equal share and marginal contribution mechanisms}
\label{subs:SAS}
We conclude this section by introducing two well-studied mechanisms that have attracted the researchers' attention due to their simple interpretation and to their special properties: the equal share mechanism (known more generally as the Shapley value) and the marginal contribution mechanism~\cite{von2013optimal}.
\begin{definition}
	The equal share and marginal contribution mechanism are identified with $\fsv$ an $\fmc$, respectively. Given $w:[n]\rightarrow \mb{R}_{>0}$, these mechanisms are defined, for $j\in[n]$, as
	\begin{align}
			\fsv(j) &= \frac{w(j)}{j}\,, \label{eq:SVdef} \\
			\fmc(j) &= w(j)-w(j-1) \label{eq:MCdef}\,,
	\end{align}		
	where we define $w(0)=0$.
\end{definition}
\noindent The equal share is the only mechanism for which the sum of the players utilities \eqref{eq:utilities} exactly match the total welfare over all instances, see \cite[Thm. 1]{part1Paccagnan2018}.
The marginal contribution mechanism takes its name from the observation \mbox{that \eqref{eq:utilities} reduces to}
\[
\begin{split}
U_i(a)&\!=\!\!\sum_{r\in a_i}v_r\fmc(|a|_r) \\
 &\!=\!\!\sum_{r\in a_i}v_r (w(|a|_r)-w(|a|_r-1))
 = W(a)-W(\emptyset,a_{-i}),
\end{split}
 \]
i.e., player's $i$ utility represents its marginal contribution to the welfare, that is the difference %
between $W(a)$ and the welfare $W(\emptyset,a_{-i})$ generated when player $i$ is removed. %

\section{The case of submodular welfare function}
\label{sec:submod}
In this section we focus on the case when the welfare basis function $w$ is non-decreasing and concave. This results in the total welfare $W(a)$ being monotone submodular. Submodular functions model problems with diminishing returns and are used to describe a wide range of engineering applications such as satellite assignment problems \cite{qu2015distributed}, Adwords for e-commerce \cite{devanur2012online}, combinatorial auctions \cite{lehmann2006combinatorial}, and many more. 

For the considered class of problems we show that the price of anarchy can be computed as the maximum between $\mc{O}(n^2)$ numbers (Theorem \ref{thm:wconcavefwdecreas}). We also give an  expression of $\poa$ for the equal share and marginal contribution mechanisms (Corollary \ref{cor:SVandMC}). We then show how to design the most efficient mechanism, and compare the resulting performance certificates with existing approximation results.
Numerical studies on the vehicle-target assignment problem conclude the section.
\begin{assumption}
\label{ass:sub}
We assume that the welfare basis $w\in\mb{R}^n_{>0}$ is non-decreasing and concave, i.e., 
\[
\begin{split}
w(j+1)&\ge w(j)\,,\\%\quad \forall j\in [n-1]\\
w(j+1)-w(j)&\le w(j)-w(j-1)\,,\quad \forall j\in [n-1]\,,
\end{split}
\]	
where we define $w(0)=0$.
\end{assumption}

\noindent As a consequence of Assumption \ref{ass:sub}, the function $W(a)$ is monotone and submodular, i.e., it satisfies the following.\\[1mm]
\emph{Monotonicity:}
\[
\forall ~ a,b\in\mc{A} \text{~s.t.~} a_i\subseteq b_i ~ \forall i\in\N,~\text{it holds}~W(a)\le W(b)\,.
\]     
\emph{Submodularity:}
\[
	\begin{split}
	&\forall ~ a,b\in\mc{A} ~\text{~s.t.~} a_i\subseteq b_i ~ \forall i\in\N,\\ %
	&\forall \,(c_1,\dots,c_n) \in {2^{\mc{R} n}} \text{~s.t.~} \tilde a_i \coloneqq a_i\!\cup c_i\!\in\mc{A}_i,~ \tilde b_i \coloneqq b_i\!\cup\! c_i \in\mc{A}_i\\ &\forall\, i\in\!\N, \text{~it holds~} W(\tilde a)-W(a)\ge W(\tilde b)- W(b)\,.
	\end{split}
\]
While Theorem \ref{thm:mine} gives a general answer on how to compute the price of anarchy through the solution of a linear program, it is possible to exploit the additional properties given by Assumption \ref{ass:sub} to obtain an explicit expression of $\poa(f,w,n)$. Towards this goal, we require the mechanism $f$ to be non-increasing and lower bounded by the corresponding marginal contribution $\fmc$, as often assumed in the literature \cite{marden2014generalized,gairing2009covering}.
\begin{theorem}[\bf $ \rm{\bf PoA}$ for submodular welfare]
\label{thm:wconcavefwdecreas}
Let the welfare basis $w$ satisfy Assumption \ref{ass:sub}, and let $f\in\mb{R}^n$.

\noindent 
$\bullet$ If $f(1)\le 0$, then $\poa(f,w,n)=0$ for any $n\in\mb{N}$.

\noindent 
$\bullet$ If instead $f(1)>0$,  $f(j)$ is non-increasing, $f(j)\ge f^{\rm mc}(j)$ for all $j\in[n]$, and Assumption \ref{ass:fw=1} holds, then $\poa(f,w,n)=1/W^\star$, where
\be
\small
\label{eq:poasubmodular}
	\begin{split}
	W^\star\!=\! \max_{
	\substack{j,l \in [n]\\[0.4mm] j\ge l}}\biggl\{&\!
	\frac{w(l)}{w(j)}\!+\!\min(j,n\!-\!l)\frac{f(j)}{w(j)}\!-\!\min(l,n\!-\!j)\frac{f(j\!+\!1)}{w(j)}\biggl
	\},
	\end{split}
\ee
and $f(n+1)=0$.
\end{theorem}

Among many others, the equal share and marginal contribution mechanisms satisfy the assumptions of Theorem \ref{thm:wconcavefwdecreas}. Thus, a direct application of Theorem \ref{thm:wconcavefwdecreas} returns the price of anarchy of $\fsv$ and $\fmc$ as the maximum between ${n(n+1)/2}$ and $n$ numbers, respectively.
\begin{corollary}[\bf Exact ${\rm {\bf PoA}}$ for $\fsv$ and $\fmc$]
Let the welfare basis $w$ satisfy Assumptions \ref{ass:fw=1} and \ref{ass:sub}.
\label{cor:SVandMC} 
\item[~~\bf i)] The price of anarchy of the equal share mechanism is $\poa(\fsv,w,n)=1/\Wsv$, where
\be
\label{eq:poafsv}
	\small
	\Wsv =\!\! \max_{
	\substack{j,l \in [n]\\[0.4mm] j\ge l}
	}\!\biggl\{\!
	\frac{w(l)}{w(j)}+\min(j,n-l)\frac{1}{j}
	-\min(l,n-j)\frac{w(j+1)}{(j+1)w(j)}\!\biggl
	\},
\ee
and $w(n+1)=0$.
\item[~~\bf ii)]
The price of anarchy for the marginal contribution mechanism is $\poa({\fmc},w,n)=1/\Wmc$, where
\be
\small
\begin{split}
\label{eq:poafmc}
		\Wmc =\!1\!+\!\max_{j \in [n]}\biggl\{\!\frac{1}{w(j)}\min(j,n\!-\!j)[2w(j)\!-\!w(j\!-\!1)\!-\!w(j\!+\!1)]\!\biggl\},
\end{split}
\ee
and $w(0)=w(n+1)=0$.
\end{corollary}
\begin{remark}
\label{rmk:comparisontim}
The result in Corollary \ref{cor:SVandMC} strengthens the findings in \cite[Thm. 6]{marden2014generalized}, showing that the bound on $\poa(\fsv,w,n)$ derived in \cite[Thm. 6]{marden2014generalized} is tight.\footnote{More in details, the result in \cite[Thm. 6]{marden2014generalized} provides an estimate of the price of anarchy relative to $\fsv$, as the minimum between two expression. While their first expression exactly matches \eqref{eq:poafsv}, the second one is not present here. 
Nevertheless, it is possible to show that such additional expression is redundant, as the first one is always the most constraining (this statement is not formally shown here, in the interest of space; its proof amounts to showing that the second expression appearing in \cite[Thm. 6]{marden2014generalized} is always upper bounded by \eqref{eq:poafsv}, thanks to the concavity of $w$).
We conclude that the bound of \cite[Thm. 6]{marden2014generalized} precisely matches the one in \eqref{eq:poafsv}. Since our result is provably tight for the class of pure Nash equilibria, and the result in  \cite{marden2014generalized} gives a lower bound for coarse correlated equilibria (CCE), such bound is tight also for the set of CCE, and the worst-performing CCE is, simply, a pure Nash equilibrium.}
\end{remark}
For the submodular welfare case considered here, it is still possible to determine the mechanism that maximizes $\poa(f,w,n)$ as the solution of a tractable linear program either directly employing the more general result in \eqref{eq:proppart2}  or using the following linear program derived from \eqref{eq:poasubmodular}, which additionally constrains the admissible mechanism $f$ to satisfy $f(j)\ge \fmc(j)$ and $f(j)$ to be non-increasing,
\be
\small
	\begin{split}
	&(\fopt,\muopt) \in \argmin_{f\in{\mc{F}},\, \mu\in\mb{R}} \mu  \\[0.1cm]
	&\,\text{s.t.}~ \,\mu w(j)\!\ge \!w(l)+j f(j) \!-\!l f(j\!+\!1) \\%1\le j+l\le k\\
	&~\hspace*{25mm} \forall j,l\in[n]~{s.t.}~ j\ge l ~~\text{and}~~ j+l\le n,\\[0.15cm]
	&\!\qquad \mu w(j)\!\ge\! w(l)\!+\!(n\!-\!l) f(j) \!-\!(n\!-\!j) f(j\!+\!1)\\
	&~\hspace*{25mm} \forall j,l\in[n] ~{s.t.}~ j\ge l ~~\text{and}~~ j+l> n\,,
	\end{split}
	\label{eq:optimalfconcave}
\ee
	where 
	$\mc{F}\!=\!\{f\!\in\mb{R}^n\,|f(1)= 1, \,f(j)\!\ge\!\fmc(j),~ f(j+1)\!\le\! f(j),~\forall j\in[n]\}$, and $f(n+1)=0$.
	\begin{remark}
	While the latter program features the same number of decision variables of that in \eqref{eq:proppart2}, the program in \eqref{eq:optimalfconcave} includes $n(n+1)/2$ constraints against $\mc{O}(n^3)$ constraints of the program in \eqref{eq:optimalfconcave}. Since the running time of  commonly employed linear program solvers (e.g., the simplex algorithm) increases with both the number of decision variables and the number of constraints, the program \eqref{eq:optimalfconcave} can be solved more efficiently than that in \eqref{eq:proppart2}.
	Note that \eqref{eq:optimalfconcave} is easily derived from the program in \eqref{eq:programreducedproof} appearing in the proof of Theorem \ref{thm:wconcavefwdecreas}, by letting $f$ be an additional decision variable.
	\end{remark}

\subsection{Comparison with existing result}
For the general class of submodular maximization problems subject to matroid constraints, the best approximation ratio achievable in polynomial time has been shown to be~\cite{sviridenko2017optimal} 
\be
1-\frac{c}{e}
\label{eq:1-ce}\,,
\ee
where $c$ represents the curvature of the welfare function (see \cite{conforti1984submodular} for its definition) and $e$ the Euler's number. 
For this class of problems, no polynomial time algorithm can do better than \eqref{eq:1-ce} on all instances. 
Relative to the class of problems considered here, i.e., those problems where $W$ has the special structure in \eqref{eq:welfaredef}, the curvature can be computed as $c=1+w(n-1)-w(n)$, yielding an approximation ratio of 
\be
{\rm App}=1-\frac{1+w(n-1)-w(n)}{e}\,.
\label{eq:approx}
\ee

Figure \ref{fig:comparisonp2} compares the approximation ratio in \eqref{eq:approx} with the price of anarchy of the equal share, marginal contribution and optimal mechanism $\fopt$, in the case when $w(j)=j^d$, $0\le d\le 1$, $|N|\le 20$. The values of the price of anarchy have been computed using \eqref{eq:poafsv}, \eqref{eq:poafmc} and \eqref{eq:proppart1} respectively, where $\fopt$ has been determined as the solution to \eqref{eq:proppart2}. Observe how $\fopt$ outperforms the approximation ratio \eqref{eq:approx}, as well as the equal share and marginal contribution mechanisms.
\begin{remark}
The fact that $\fopt$ gives performance guarantees beyond \eqref{eq:approx} is not in contradiction with the inapproximability result presented in \cite{sviridenko2017optimal}, as the welfare in \eqref{eq:welfaredef} has \emph{special form}. 
\end{remark}
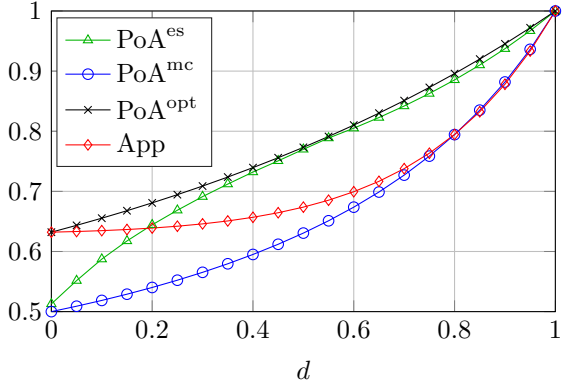
\begin{figure}[ht!] 
\begin{center}
\hspace*{-5mm}
\newlength\figureheight 
\newlength\figurewidth 
\setlength\figureheight{4cm} 
\setlength\figurewidth{6.7cm} 
\begin{tikzpicture}

\begin{axis}[%
width=\figurewidth,
height=\figureheight,,
at={(1.011in,0.642in)},
scale only axis,
xmin=0,
xmax=1,
xlabel={$d$},
ymin=0.5,
ymax=1,
ytick={0.5, 0.6, 0.7, 0.8, 0.9 ,1},
grid=both,
legend style={at={(0.01,0.73)}, anchor=west, draw=white!15!black},
legend cell align={left}
]

\addplot[color=green,mark=triangle,mark options={solid}]
  table[row sep=crcr]{%
0	0.512820512820513\\
0.05	0.551579892207097\\
0.1	0.587291533788722\\
0.15	0.617541516463445\\
0.2	0.644361679553719\\
0.25	0.668613433410656\\
0.3	0.691233805537062\\
0.35	0.71209950483051\\
0.4	0.732154497890012\\
0.45	0.750757656686259\\
0.5	0.76990681212978\\
0.55	0.788670183689739\\
0.6	0.805061118646457\\
0.65	0.822763630350338\\
0.7	0.841929713797954\\
0.75	0.862735663283182\\
0.8	0.885387134211467\\
0.85	0.910125525400614\\
0.9	0.937236102023765\\
0.95	0.967058439409503\\
1	1\\
};
\addlegendentry{$\poa^{\rm es}$};
\addplot [color=blue,mark=o,mark options={solid}]
  table[row sep=crcr]{%
0	0.5\\
0.05	0.508974473014022\\
0.1	0.518611263028954\\
0.15	0.528980031386739\\
0.2	0.540160487999998\\
0.25	0.55224427281837\\
0.3	0.565337276567052\\
0.35	0.579562525251956\\
0.4	0.595063794950633\\
0.45	0.612010182095219\\
0.5	0.630601937481871\\
0.55	0.651077991369887\\
0.6	0.673725770588819\\
0.65	0.698894165659515\\
0.7	0.727010893750765\\
0.75	0.758606100120772\\
0.8	0.794344979667087\\
0.85	0.835073714094604\\
0.9	0.881885528675272\\
0.95	0.936217958793979\\
1	1\\
};
\addlegendentry{$\poa^{\rm mc}$\hspace*{-1mm}};
%\addplot [color=red,mark=square,mark options={solid}]
%  table[row sep=crcr]{%
%%
%0	0.632120558828558\\
%0.05	0.63290745689283\\
%0.1	0.633948500190576\\
%0.15	0.635306399990088\\
%0.2	0.637058289861445\\
%0.25	0.639299182932335\\
%0.3	0.642146390943544\\
%0.35	0.645745237489006\\
%0.4	0.650276538858866\\
%0.45	0.655966537117448\\
%0.5	0.663100291363061\\
%0.55	0.672040030296724\\
%0.6	0.683250752894955\\
%0.65	0.697336624484692\\
%0.7	0.715093788519244\\
%0.75	0.73758870772346\\
%0.8	0.766277195099631\\
%0.85	0.803190075166655\\
%0.9	0.851231268101469\\
%0.95	0.914671985458154\\
%0.9999	0.999801547920889\\
%};
%\addlegendentry{$\fmc$};

\addplot[color=black,solid,mark=x,mark options={solid}]
  table[row sep=crcr]{%
0	0.632120555132191\\
0.05	0.643482913435981\\
0.1	0.655375384527601\\
0.15	0.667825056206542\\
0.2	0.680860546884625\\
0.25	0.694512001740175\\
0.3	0.708811156686717\\
0.35	0.723791477212326\\
0.4	0.739487634182099\\
0.45	0.755936488911006\\
0.5	0.773180596528451\\
0.55	0.791262194488063\\
0.6	0.810212760997551\\
0.65	0.830092814964609\\
0.7	0.850955470654624\\
0.75	0.872854275447214\\
0.8	0.895845961995925\\
0.85	0.919990663553415\\
0.9	0.945352121634846\\
0.95	0.971997956707326\\
1	1\\
};
\addlegendentry{$\poa^{\rm opt}$};

\addplot[color=red,solid,mark=diamond,mark options={solid}]
  table[row sep=crcr]{%
0      0.632120558828558\\
0.05   0.633215096779674\\
0.10   0.634660102977066\\
0.15   0.636539744118561\\
0.20   0.638956154576726\\
0.25   0.642033032461418\\
0.30   0.645919926631880\\
0.35   0.650797343733750\\
0.40   0.656882827957390\\
0.45   0.664438194090168\\
0.50   0.673778127323984\\
0.55   0.685280402070890\\
0.60   0.699398017781229\\
0.65   0.716673603681631\\
0.70   0.737756507901600\\
0.75   0.763423061338450\\
0.80   0.794600594819830\\
0.85   0.832395892009683\\
0.90   0.878128882826179\\
0.95   0.933372526139581\\
1.00   1.000000000000000\\
};
\addlegendentry{${\rm App}$};

\end{axis}
\end{tikzpicture}% 
\caption{
Comparison between the  approximation ratio \eqref{eq:approx} and the price of anarchy of the optimal mechanism $\fopt$ (determined as the solution of \eqref{eq:proppart2}), equal share $\fsv$ and marginal contribution $\fmc$ mechanisms, denoted with  $\poa^{\rm opt}$, $\poa^{\rm es}$ and $\poa^{\rm mc}$, respectively.
The problems considered feature $|N|\le 20$ and $w(j)=j^d$ with $0\le d\le 1$ represented over the $x$-axis.
First, we observe that the optimal mechanism $\fopt$ outperforms \eqref{eq:approx} for all values of $d$, so that, when there exists an algorithm capable of computing a pure Nash equilibrium in polynomial time (see Proposition \ref{prop:poly}), the approach presented here gives improved guarantees compared to \eqref{eq:1-ce}.
Second, we note that the equal share mechanism performs close to the optimal for values of $0.5\le d\le 1$, while its performance degrades for $0\le d< 0.5$ reaching its lowest value of $1/2$ for $d=0$, as predicted in \cite[Thm. 1]{part1Paccagnan2018}. The marginal contribution mechanism instead, performs the worst amongst the considered ones. While $\fopt$ will always perform better or equal than any other mechanism, it is unclear if, and to what extent, $\fsv$ outperforms $\fmc$ in the general settings. The expressions in \eqref{eq:poafsv} and \eqref{eq:poafmc} could nevertheless be used to provide an answer to this question.
} 
\label{fig:comparisonp2}
\end{center}
\end{figure}

\subsection{Application: the vehicle-target assignment problem}
\label{subsec:vehicletarget}
In the following we consider the vehicle-target assignment problem introduced in \cite{murphey2000target} and studied in, e.g., \cite{arslan2007autonomous, marden2014generalized}. 
Let $\mc{R}$ be a finite set of targets, and for each target $r\in\mc{R}$ let $v_r\ge 0$ represent its relative importance. Let $N$ be a set of vehicles, and for each vehicle let $\mc{A}_i\in 2^{\mc{R}}$ be a set of feasible target assignments. The goal is to distributedly compute a feasible allocation $a\in\mc{A}$ so as to maximize the joint probability of successfully acquiring the targets, expressed as 
\[
W(a)\coloneqq\sum_{r\in\cup_{i\in N} a_i}v_r (1-(1-p)^{|a|_r}),
\]
where $(1-(1-p)^{|a|_r})$ is the probability that $|a|_r$ vehicles eliminate the target $r$, while the scalar quantity $0<p\le 1$ is a parameter representing the probability that a vehicle will successfully acquire a target. In the forthcoming presentation, it is assumed that the success probability $p$ is the same for all vehicles. Observe that the welfare considered here has the form \eqref{eq:welfaredef} with welfare basis $1-(1-p)^{|a|_r}$. We normalize this quantity (without affecting the problem's solution) so that $w(1)=1$ and thus define 
\be
w(j)\coloneqq \frac{1-(1-p)^{j}}{p}.
\label{eq:wassignment}
\ee
Observe that \eqref{eq:wassignment} satisfies Assumption \ref{ass:sub} in that $w(j)>0$ and $w(j)$ is increasing and concave. Thus, it is possible to compute the performance guarantee of any set of utility functions of the form \eqref{eq:utilities}, and to determine the best mechanism $f\in\mb{R}^n$ by solving a corresponding linear program. 

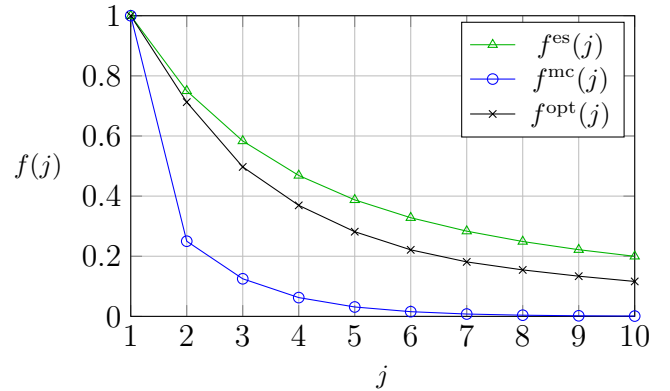
\begin{figure}[ht!] 
\begin{center}
\hspace*{-5mm} 
\setlength\figureheight{4cm} 
\setlength\figurewidth{6.7cm} 
% Minimal pgfplots version: 1.3
%  http://www.mathworks.com/matlabcentral/fileexchange/22022-matlab2tikz
\definecolor{mycolor1}{rgb}{0.85000,0.32500,0.09800}%
\begin{tikzpicture}

\begin{axis}[%
width=\figurewidth,
height=\figureheight,
at={(1.011111in,0.641667in)},
scale only axis,
xmin=1,
xmax=10,
xtick={1,2,3,4,5,6,7,8,9,10},
xlabel={$j$},
ymin=0,
ymax=1,
ylabel={$f(j)$},
ylabel style = {rotate=-90},
grid = both,
ticklabel style = {font=\large},
]

\addplot [color=green,mark=triangle,mark options={solid}]
  table[row sep=crcr]{%
1 1.00000000000\\
2 0.75000000000\\
3 0.58333333333\\
4 0.46875000000\\
5 0.38750000000\\
6 0.32812500000\\
7 0.28348214286\\
8 0.24902343750\\
9 0.22178819444\\
10 0.19980468750\\
};
%\addlegendentry{~$\fsv(j)/\fsv(1)$};
\addlegendentry{~$\fsv(j)$};
\addplot [color=blue,mark=o,mark options={solid}]
  table[row sep=crcr]{%
1 1.0000000000\\
2 0.2500000000\\
3 0.1250000000\\
4 0.0625000000\\
5 0.0312500000\\
6 0.0156250000\\
7 0.0078125000\\
8 0.0039062500\\
9 0.0019531250\\
10 0.0009765625\\
};
%\addlegendentry{~$\fmc(j)/\fmc(1)$};
\addlegendentry{~$\fmc(j)$};
\addplot [color=black,solid,mark=x,mark options={solid}]
  table[row sep=crcr]{%
1 1\\
2 0.712649086\\
3 0.497135901\\
4 0.369271802\\
5 0.281652122\\
6 0.221339404\\
7 0.181188105\\
8 0.154576589\\
9 0.133695861\\
10 0.116197912\\};
%\addlegendentry{~$\fopt(j)/\fopt(1)$};
\addlegendentry{~$\fopt(j)$};
\end{axis}
\end{tikzpicture}% 
\caption{Mechanisms $\fsv$, $\fmc$ and optimal mechanisms $\fopt$ obtained solving \eqref{eq:proppart2} for the specific choice of $w$ in \eqref{eq:wassignment} with $|N|\le 10$ and $p=0.5$. 
Note how $\fopt$ is non-increasing and naturally satisfies $\fopt(j)\ge \fmc(j)$ even if these constraints have not been a priori enforced.} 
\label{fig:optf_assignment}
\end{center}
\end{figure}

Figure \ref{fig:comparison_sensoralloc} shows the achievable approximation ratios for the equal share, marginal contribution, optimal mechanisms, as well as the bound in \eqref{eq:approx}. Note how the optimal mechanism significantly outperforms all the others as well as the bound of \eqref{eq:approx} for non-trivial values of $p$. 
Figure \ref{fig:optf_assignment} shows the mechanisms $\fsv$, $\fmc$ and $\fopt$,  highlighting the fact that $\fopt$ naturally satisfies $\fopt(j)\le \fopt(j+1)$ and $\fopt(j)\ge \fmc(j)$ even if these constraints have not been a priori enforced.
\begin{figure}[ht!] 
\begin{center}
\hspace*{-5mm} 
\setlength\figureheight{4cm} 
\setlength\figurewidth{6.7cm} 
\begin{tikzpicture}

\begin{axis}[%
width=\figurewidth,
height=\figureheight,,
at={(1.011in,0.642in)},
scale only axis,
xmin=0,
xmax=1,
xlabel={$p$},
ymin=0.5,
ymax=1,
ytick={0.5, 0.6, 0.7, 0.8, 0.9 ,1},
grid=both,
legend style={at={(0.62,0.73)}, anchor=west, draw=white!15!black},
legend cell align={left}
]
% 10 AGENTS
\addplot[color=green,mark=triangle,mark options={solid}]
  table[row sep=crcr]{%
0.01 0.987595526822133\\
0.05	0.94005506248334\\
0.1	0.885945929385683\\
0.15	0.836703243672895\\
0.2	0.792757768782388\\
0.25	0.755066591317223\\
0.3	0.723594624162156\\
0.35	0.696731173009963\\
0.4	0.671695703060633\\
0.45	0.651151084727883\\
0.5	0.634576018857391\\
0.55	0.621432749556367\\
0.6	0.609723347378961\\
0.65	0.596116584115093\\
0.7	0.584793483986921\\
0.75	0.575539272188375\\
0.8	0.568181786446279\\
0.85	0.562587902575993\\
0.9	0.555555555305556\\
0.95	0.540540540540053\\
1	0.526315789473684\\
};
\addlegendentry{$\poa^{\rm es}$};
\addplot [color=blue,mark=o,mark options={solid}]
  table[row sep=crcr]{%
0.01    0.990099009900991\\
0.05	0.952380952380951\\
0.1	0.909090909090909\\
0.15	0.869565217391305\\
0.2	0.833333333333333\\
0.25	0.8\\
0.3	0.769230769230769\\
0.35	0.740740740740741\\
0.4	0.714285714285714\\
0.45	0.689655172413793\\
0.5	0.666666666666667\\
0.55	0.645161290322581\\
0.6	0.625\\
0.65	0.606060606060606\\
0.7	0.588235294117647\\
0.75	0.571428571428571\\
0.8	0.555555555555556\\
0.85	0.54054054054054\\
0.9	0.526315789473684\\
0.95	0.512820512820513\\
1	0.5\\
};
\addlegendentry{$\poa^{\rm mc}$\hspace*{-1mm}};

\addplot[color=black,solid,mark=x,mark options={solid}]
  table[row sep=crcr]{%
0.01  0.992953665000470\\
0.05	0.966389234043331\\
0.1	0.936548216525508\\
0.15	0.909899317621329\\
0.2	0.885696582179183\\
0.25	0.8635731105029\\
0.3	0.84338976834284\\
0.35	0.824982401255014\\
0.4	0.807881209929548\\
0.45	0.791907322734352\\
0.5	0.776788977492372\\
0.55	0.76108086993257\\
0.6	0.745541787689185\\
0.65	0.730655009452087\\
0.7	0.71639163972289\\
0.75	0.702577389606777\\
0.8	0.687968462507212\\
0.85	0.673586478787689\\
0.9	0.65936662214631\\
0.95	0.645535765279937\\
1	0.632120559276055\\
};
\addlegendentry{$\poa^{\rm opt}$};

\addplot[color=red,solid,mark=diamond,mark options={solid}]
  table[row sep=crcr]{%
  0.01 0.968184773333313\\
0.05	0.863976359476678\\
0.1	0.774644591820245\\
0.15	0.717327671593058\\
0.2	0.681496501600498\\
0.25	0.659742671257519\\
0.3	0.64696582122097\\
0.35	0.639740045748953\\
0.4	0.635827936001333\\
0.45	0.633814778513864\\
0.5	0.632839073362096\\
0.55	0.632398926080495\\
0.6	0.632216996216784\\
0.65	0.632149553481668\\
0.7	0.632127799799598\\
0.75	0.632121962177256\\
0.8	0.632120747182832\\
0.85	0.632120572971079\\
0.9	0.632120559196437\\
0.95	0.632120558829276\\
1	0.632120558828558\\
};
\addlegendentry{${\rm App}$};

\end{axis}
\end{tikzpicture}% 
\caption{
Comparison between the approximation ratio \eqref{eq:approx} and the price of anarchy of the optimal mechanism $\fopt$, the equal share $\fsv$ and marginal contribution $\fmc$ mechanisms, denoted with $\poa^{\rm opt}$, $\poa^{\rm es}$ and $\poa^{\rm mc}$, respectively.
The problems considered features $|N|\le 10$ vehicles and welfare basis as in \eqref{eq:wassignment} with $0<p\le 1$ represented over the $x$-axis. Note how the optimal mechanism significantly outperforms all the others as well as the bound of \eqref{eq:approx} for non-trivial values of $p$. 
In the extreme case of $p=1$, $\fopt$ matches \eqref{eq:approx}, while for small $p$ all the design methodologies offer a similarly high performance guarantee.
}
\label{fig:comparison_sensoralloc}
\end{center}
\end{figure}

In both Figures \ref{fig:comparison_sensoralloc} and \ref{fig:optf_assignment} we have set the number of agents to be relatively small\footnote{{Similar trends and conclusions can be obtained with larger values of $n$.}}, i.e., $|N|\le 10$. This choice was purely made so as to perform an exhaustive search simulation in order to test the provided bounds displayed in Figure \ref{fig:comparison_sensoralloc}. More specifically, we considered $10^5$ random instances of the vehicle target assignment problem, constructed as follows. Each instance features $n=10$ agents, $n+1$ resources and fixed $p=0.8$. Each agent is equipped with an action set with only two allocations. Each of the two allocations consists of a single resource, uniformly randomly chosen from the $n+1$ available. 
We believe this is not restrictive in assessing the performance, as the structure of some worst case instances is of this form \cite{paccagnan2017arxiv}.
The values of the resources are randomly generated with uniform distribution in the interval $0\le v_r\le 1$.
{Observe that any constraint set $\mc{A}_i$ is the basis of a uniform matroid of rank one (upon adding the allocation $\emptyset$ to each $\mc{A}_i$), see Appendix \ref{appendixa} for a definition. Further note that computing a single best response is a polynomial operation in the number of resources. Thus, it is possible to compute a pure Nash equilibrium in polynomial time, as claimed in Proposition \ref{prop:poly} included in the forthcoming Section \ref{sec:complexity}. It follows that the guarantees offered by $\poa$ can be achieved quickly.

 For this class of problems, the theoretical worst case performance is $\poa(\fsv,w,n) \approx 0.568 $, $\poa(\fmc,w,n)\approx  0.556$, $\poa(\fopt,w,n)\approx 0.688$ (see Figure \ref{fig:comparison_sensoralloc} with $p=0.8$).
For each instance $G$ generated, we performed an exhaustive search so as to compute the welfare at the worst equilibrium $\min_{a\in\nashe{G}}W(a)$ and the value $W(\aopt)$. The ratio between these quantities (their empirical cumulative distribution) is plotted across the $10^5$ samples in Figure \ref{fig:CDF_comparison}, for $\fsv$, $\fmc$, $\fopt$. In the same figure the vertical dashed lines represent the theoretical bounds on the price of anarchy, while the markers represent the worst case performance occurred in simulations.
A table comparing the worst case performance encountered in the simulations and the theoretical values of the price of anarchy is also included.
\begin{figure}[ht!] 
\begin{center}
{\small 
\hspace*{6mm}
\begin{tabular}{ccc}
{Mechanism} & {$\poa(f,w,n)$} & {Empirical $\poa$} \\ 
\specialrule{.05em}{.2em}{.2em} 
$\fsv$ & $0.568$ & $0.744$ \\
\specialrule{.05em}{.2em}{.2em} 
$\fmc$ & $0.556$ & $0.715$\\
\specialrule{.05em}{.2em}{.2em} 
$\fopt$ & $0.688$ & $0.802$\\
\specialrule{.05em}{.2em}{.2em} 
\vspace*{2mm} 
\end{tabular}}\\
\hspace*{-10mm} 
\setlength\figureheight{4cm} 
\setlength\figurewidth{6.7cm} 
\begin{tikzpicture}

\begin{axis}[%
width=\figurewidth,
height=\figureheight,,
at={(1.011in,0.642in)},
scale only axis,
xmin=0.54,
xmax=1,
xlabel={$\frac{\min_{a\in\nashe{G}}W(a)}{W(\aopt)}$},
ymin=-0.05,
ymax=1,
unbounded coords=jump,
ytick={0, 0.25, 0.5, 0.75, 1},
ylabel={CDF\hspace*{-2mm}},
ylabel style={rotate=-90},
grid=both,
legend style={at={(0.68,0.76)}, anchor=west, draw=white!15!black},
legend cell align={left}
]
% vertical lines
\addplot[color=green, dashed , line width = 0.5pt, forget plot]
table[row sep=crcr]
  {0.568 	-0.1\\
   0.568 	1\\};
   \addplot[color=blue, dashed , line width = 0.5pt, forget plot]
table[row sep=crcr]
  {0.556 	-0.1\\
   0.556 	1\\};
   \addplot[color=black, dashed , line width = 0.5pt, forget plot]
table[row sep=crcr]
  {0.688 	-0.1\\
   0.688 	1\\};
% Plots of single point to get desired legend
%
\addplot[color=green, mark = triangle, mark size = 4pt, mark options={solid}, line width = 1pt]
  table[row sep=crcr]
  {0.74459	1.25e-05\\};
\addlegendentry{$\fsv$\hspace*{-1mm}};
\addplot [color=blue, mark = o, mark size = 3pt, line width = 1pt]
  table[row sep=crcr]
  {%
0.71459	2.5e-05\\};
\addlegendentry{$\fmc$\hspace*{-1mm}};
\addplot[color=black, mark = x, mark size = 4pt, line width = 1pt]
  table[row sep=crcr]{%
0.802	3.75e-05\\};
\addlegendentry{$\fopt$};

%%%% real plots

\addplot [color=blue, line width = 1pt, forget plot]
  table[row sep=crcr]
  {%
0.71459	2.5e-05\\
0.74718	2.5e-05\\
0.74977	2.5e-05\\
0.75236	2.5e-05\\
0.75495	2.5e-05\\
0.75754	2.5e-05\\
0.76013	2.5e-05\\
0.76272	2.5e-05\\
0.76531	3.75e-05\\
0.7679	6.25e-05\\
0.77049	7.5e-05\\
0.77308	8.75e-05\\
0.77567	0.0001\\
0.77826	0.0001\\
0.78085	0.0001125\\
0.78344	0.000175\\
0.78603	0.000225\\
0.78862	0.0002625\\
0.79121	0.000325\\
0.7938	0.000375\\
0.79639	0.000425\\
0.79898	0.0004625\\
0.80157	0.0005\\
0.80416	0.0006125\\
0.80675	0.0007\\
0.80934	0.000875\\
0.81193	0.0009625\\
0.81452	0.0010625\\
0.81711	0.001225\\
0.8197	0.0014625\\
0.82229	0.001625\\
0.82488	0.001875\\
0.82747	0.0021\\
0.83006	0.0024875\\
0.83265	0.0029\\
0.83524	0.0034125\\
0.83783	0.0039375\\
0.84042	0.00455\\
0.84301	0.0050375\\
0.8456	0.0058\\
0.84819	0.006675\\
0.85078	0.0078125\\
0.85337	0.00915\\
0.85596	0.01055\\
0.85855	0.011775\\
0.86114	0.0136625\\
0.86373	0.0153125\\
0.86632	0.0172375\\
0.86891	0.0192875\\
0.8715	0.0215125\\
0.87409	0.0242\\
0.87668	0.0273875\\
0.87927	0.0306\\
0.88186	0.0342\\
0.88445	0.0379625\\
0.88704	0.0424\\
0.88963	0.0465625\\
0.89222	0.0516625\\
0.89481	0.0563375\\
0.8974	0.0619625\\
0.89999	0.067875\\
0.90258	0.0747875\\
0.90517	0.082175\\
0.90776	0.0900375\\
0.91035	0.0976\\
0.91294	0.1063875\\
0.91553	0.11485\\
0.91812	0.1245875\\
0.92071	0.1340625\\
0.9233	0.1443875\\
0.92589	0.1547125\\
0.92848	0.1658\\
0.93107	0.177275\\
0.93366	0.1895125\\
0.93625	0.2012125\\
0.93884	0.213525\\
0.94143	0.2259875\\
0.94402	0.23905\\
0.94661	0.2526625\\
0.9492	0.2658375\\
0.95179	0.2797\\
0.95438	0.29335\\
0.95697	0.3081875\\
0.95956	0.3231875\\
0.96215	0.3397\\
0.96474	0.3573875\\
0.96733	0.3775625\\
0.96992	0.400475\\
0.97251	0.4269375\\
0.9751	0.4575625\\
0.97769	0.493025\\
0.98028	0.5322125\\
0.98287	0.575875\\
0.98546	0.6218625\\
0.98805	0.670175\\
0.99064	0.7180875\\
0.99323	0.765075\\
0.99582	0.8113375\\
0.99841	0.853225\\
1.001	1\\
};

\addplot[color=green,mark options={solid}, line width = 1pt, forget plot]
  table[row sep=crcr]
  {%
0.74459	1.25e-05\\
0.80008	1.25e-05\\
0.80212	3.75e-05\\
0.80416	3.75e-05\\
0.8062	6.25e-05\\
0.80824	0.0001125\\
0.81028	0.0001375\\
0.81232	0.00015\\
0.81436	0.0001875\\
0.8164	0.0002375\\
0.81844	0.000275\\
0.82048	0.00035\\
0.82252	0.00045\\
0.82456	0.0005125\\
0.8266	0.0006375\\
0.82864	0.0008125\\
0.83068	0.0009625\\
0.83272	0.00115\\
0.83476	0.0012875\\
0.8368	0.001475\\
0.83884	0.00165\\
0.84088	0.0019125\\
0.84292	0.002275\\
0.84496	0.002675\\
0.847	0.003125\\
0.84904	0.0036125\\
0.85108	0.0041625\\
0.85312	0.0048625\\
0.85516	0.00545\\
0.8572	0.0060875\\
0.85924	0.007025\\
0.86128	0.0080125\\
0.86332	0.00915\\
0.86536	0.0102375\\
0.8674	0.0114375\\
0.86944	0.0127125\\
0.87148	0.0141875\\
0.87352	0.015875\\
0.87556	0.0175625\\
0.8776	0.019575\\
0.87964	0.0217625\\
0.88168	0.0241875\\
0.88372	0.0266\\
0.88576	0.029375\\
0.8878	0.032625\\
0.88984	0.03595\\
0.89188	0.0397125\\
0.89392	0.0439125\\
0.89596	0.048225\\
0.898	0.0526375\\
0.90004	0.05815\\
0.90208	0.0635\\
0.90412	0.0689625\\
0.90616	0.0755\\
0.9082	0.0820125\\
0.91024	0.0895375\\
0.91228	0.097525\\
0.91432	0.1049125\\
0.91636	0.1134625\\
0.9184	0.1226125\\
0.92044	0.1315375\\
0.92248	0.1414125\\
0.92452	0.1520875\\
0.92656	0.1637875\\
0.9286	0.1751125\\
0.93064	0.187525\\
0.93268	0.200175\\
0.93472	0.213325\\
0.93676	0.2272625\\
0.9388	0.2422125\\
0.94084	0.2589\\
0.94288	0.27505\\
0.94492	0.291975\\
0.94696	0.309425\\
0.949	0.3276125\\
0.95104	0.34575\\
0.95308	0.364675\\
0.95512	0.384075\\
0.95716	0.4046125\\
0.9592	0.424275\\
0.96124	0.4445125\\
0.96328	0.4653\\
0.96532	0.4869875\\
0.96736	0.5073125\\
0.9694	0.528825\\
0.97144	0.5506875\\
0.97348	0.572625\\
0.97552	0.5948875\\
0.97756	0.6176625\\
0.9796	0.6416875\\
0.98164	0.6652\\
0.98368	0.6896125\\
0.98572	0.7141375\\
0.98776	0.7370625\\
0.9898	0.7595625\\
0.99184	0.781275\\
0.99388	0.80335\\
0.99592	0.825325\\
0.99796	0.848475\\
1	1\\
};
\addplot[color=black, line width = 1pt, forget plot]
  table[row sep=crcr]{%
0.802	3.75e-05\\
0.804	3.75e-05\\
0.806	3.75e-05\\
0.808	5e-05\\
0.81	6.25e-05\\
0.812	6.25e-05\\
0.814	6.25e-05\\
0.816	7.5e-05\\
0.818	7.5e-05\\
0.82	8.75e-05\\
0.822	0.0001125\\
0.824	0.000125\\
0.826	0.00015\\
0.828	0.0001875\\
0.83	0.00025\\
0.832	0.0003\\
0.834	0.000375\\
0.836	0.0004125\\
0.838	0.0005\\
0.84	0.0005625\\
0.842	0.0007125\\
0.844	0.0008875\\
0.846	0.0010625\\
0.848	0.001225\\
0.85	0.0013625\\
0.852	0.0016125\\
0.854	0.001975\\
0.856	0.00225\\
0.858	0.0026875\\
0.86	0.003275\\
0.862	0.0037625\\
0.864	0.00435\\
0.866	0.005\\
0.868	0.0057875\\
0.87	0.00645\\
0.872	0.007375\\
0.874	0.0083125\\
0.876	0.0093375\\
0.878	0.0104125\\
0.88	0.0116375\\
0.882	0.013\\
0.884	0.0146\\
0.886	0.0162\\
0.888	0.0180375\\
0.89	0.0203625\\
0.892	0.0227375\\
0.894	0.025225\\
0.896	0.0280625\\
0.898	0.0312\\
0.9	0.0347\\
0.902	0.038425\\
0.904	0.0422375\\
0.906	0.04655\\
0.908	0.0511875\\
0.91	0.056275\\
0.912	0.062225\\
0.914	0.0674875\\
0.916	0.0734625\\
0.918	0.080025\\
0.92	0.08695\\
0.922	0.0945125\\
0.924	0.1031\\
0.926	0.112\\
0.928	0.1215875\\
0.93	0.1314875\\
0.932	0.1424125\\
0.934	0.1535125\\
0.936	0.16525\\
0.938	0.1771625\\
0.94	0.1905625\\
0.942	0.2050125\\
0.944	0.2193625\\
0.946	0.23455\\
0.948	0.250725\\
0.95	0.267775\\
0.952	0.2849\\
0.954	0.3032625\\
0.956	0.32275\\
0.958	0.34365\\
0.96	0.36425\\
0.962	0.3854375\\
0.964	0.406925\\
0.966	0.4290375\\
0.968	0.4509375\\
0.97	0.47375\\
0.972	0.4972\\
0.974	0.519975\\
0.976	0.5430375\\
0.978	0.568075\\
0.98	0.594075\\
0.982	0.6195125\\
0.984	0.6453875\\
0.986	0.67195\\
0.988	0.6982375\\
0.99	0.7241625\\
0.992	0.74945\\
0.994	0.7751125\\
0.996	0.8014125\\
0.998	0.827725\\
1	1\\
};
\end{axis}
%\node at (2, 4) {\small $\poa(\fsv)$};
%\node at (3, 4) {\small $\poa(\fmc)$};
\draw[-stealth, line width=0.5pt] (5.2 ,2.1) -- (4.8, 2.1);
\node at (6.3, 2.1) {\footnotesize $\poa(\fopt,w,n)$};
\node at (4.42, 2.5) {\footnotesize $\poa(\fmc,w,n)$};
\draw[-stealth, line width=0.5pt] (3.25 ,2.5) -- (2.85, 2.5);
\node at (4.43, 3) {\footnotesize $\poa(\fsv,w,n)$};
\draw[-stealth, line width=0.5pt] (3.25 ,3) -- (3.05, 3);
\end{tikzpicture}%
\caption{Top: comparison between the theoretical values of the price of anarchy and the ratio $\min_{a\in\nashe{G}}W(a)/W(\aopt)$ for 
$\fsv$, $\fmc$, $\fopt$ across $10^5$ samples constructed as detailed above.
Bottom: cumulative distribution of the ratio $\min_{a\in\nashe{G}}W(a)/W(\aopt)$ for 
$\fsv$, $\fmc$, $\fopt$ across the same samples. The dashed lines represent the theoretical values of $\poa(\fsv,w,n)$, $\poa(\fmc,w,n)$, $\poa(\fopt,w,n)$, while the corresponding markers identify the worst case performance encountered during the simulations.
} 
\label{fig:CDF_comparison}
\end{center}
\end{figure}

First, we observe that no instance has performed worse than the corresponding price of anarchy, as predicted by Theorem \ref{thm:mine}. Second, we note that the worst case performance encountered in the simulations is $10\%/20\%$ better than the analytical worst case, partially due to the restrictions we imposed on the structure of the instances.%
\footnote{Recall indeed, that our result in Theorem \ref{thm:mine} is tight, i.e., there exists at least one instance achieving exactly an efficiency equal to the price of anarchy.}
Further, the optimal mechanism $\fopt$ has outperformed the others also in the simulations. Its worst case performance is indeed superior to the others (markers in Figure \ref{fig:CDF_comparison}). Additionally, the cumulative distribution of $\fopt$ lies below the cumulative distributions of $\fsv$ and $\fmc$ (for abscissas smaller than $0.95$). This means that, for any given approximation ratio $0\le r\le 0.95$, there is a smaller fraction of problems on which $\fopt$ performs worse or equal to $r$, compared to $\fsv$ and $\fmc$. Observe that this is not obvious a priori, as $\fopt$ is designed to maximize the worst case performance and not, e.g., the average performance. %

\section{Weighted maximum coverage problems}
\label{sec:coveringg}
In this section we specialize the previous results to the case of weighted maximum coverage problems. 
In a weighted maximum coverage problem we are given a set of weighted resources $\mc{R}$, and a common collection $\bar{\mc{A}}\subset 2^{\mc R}$. The goal is to select $n$ subsets from $\bar{\mc{A}}$ to maximize the total weight of covered elements.
This corresponds to the case where all agents have identical allocation sets $\mc{A}_i=\bar{\mc{A}}$, and welfare
\[
W(a)=\sum_{r\in\cup_{i\in N} a_i}\hspace*{-2mm}v_r\,,
\]
which is obtained with the choice of $w=\ones[n]$ in \eqref{eq:welfaredef} and \eqref{eq:utilities}. 
In this section we consider the more general setup, sometimes referred to as the \emph{general weighted maximum coverage} problem \cite{gairing2009covering}, where the allocation sets $\mc{A}_i$ need not be the same across all agents. 
General weighted maximum coverage problems satisfy Assumption \ref{ass:sub}, and are used to model a broad-spectrum of engineering problems such as sensor allocation problems \cite{marden2013distributed}, job scheduling, facility locations \cite{Hochbaum96}. %

Relative to this class of problems, we provide a tight expression for the price of anarchy (Theorem \ref{thm:poageneralcovering}) and show how this reduces to the results obtained in \cite{gairing2009covering, paccagnan2017arxiv}, under the additional assumptions therein required. We conclude the section with an application to caching in wireless networks.

\begin{theorem}
\label{thm:poageneralcovering}
Consider $w=\ones[n]$. Let $f\in\mb{R}^n_{\ge0}$ satisfy Assumption \ref{ass:fw=1}. The price of anarchy is
$\poa(f,\ones[n],n)=1/ W^\star$,
\be
\label{eq:Wstarsetcoveringg}
W^\star = 1+\!\!\max_{j\in [n-1]}\!\{(j+1)f(j+1)-1, jf(j)-f(j+1),jf(j+1)\}.
\ee
\end{theorem}
The next Corollary shows how the result in the previous theorem matches the results in \cite{gairing2009covering, paccagnan2017arxiv}, under the additional assumptions therein required.

\begin{corollary}
\label{cor:backtogair}
Let $f\in\mb{R}^n_{\ge0}$ satisfy Assumption \ref{ass:fw=1}, and additionally be non-increasing.
The expression in \eqref{eq:Wstarsetcoveringg} reduces to
\be
\label{eq:Wstarcoveringgair}
W^\star = 1 +\max_{j\in [n-1]}\{jf(j)-f(j+1),(n-1)f(n)\}\,.
\ee	
\end{corollary}
\begin{remark}
Theorem \ref{thm:poageneralcovering} extends the previous bounds derived in \cite{gairing2009covering, paccagnan2017arxiv}. In the latter works, the authors additionally required the admissible mechanisms to be non-increasing and such that $jf(j)\le 1$ for all $j\in [n]$.
More precisely, in \cite[Thm. 2]{gairing2009covering} the author provides a bound matching the expression in \eqref{eq:Wstarcoveringgair}, while tightness of the previous bound is shown in \cite[Thm. 1]{paccagnan2017arxiv}.  
\end{remark}

While \cite[Eq. 5]{gairing2009covering} determines the mechanism maximizing the price of anarchy \eqref{eq:Wstarcoveringgair} as
\be
\fopt(j) \coloneqq(j-1)!\frac
{\frac{1}{(n-1)(n-1)!} +\sum_{i=j}^{n-1}\frac{1}{i!}}
{\frac{1}{(n-1)(n-1)!} +\sum_{i=1}^{n-1}\frac{1}{i!}},\quad j\in[n]\,,
\label{eq:fstargair}
\ee
 the set of feasible mechanisms in \cite{gairing2009covering} is further limited to $jf(j)\le 1$ and $f$ non-increasing.
Using the result provided in Theorem \ref{thm:poageneralcovering} it is possible to determine the best mechanism (via a linear program derived from \eqref{eq:Wstarsetcoveringg}) without imposing these additional constraints. 
Numerical simulations show that the optimal mechanism obtained optimizing \eqref{eq:Wstarsetcoveringg} matches that in \eqref{eq:fstargair}, so that removing the additional assumption therein required does not improve the optimal price of anarchy.\footnote{This statement can be formally proved, by showing that the mechanism in \eqref{eq:fstargair} solves the KKT system corresponding to the problem of minimizing $W^\star$ in \eqref{eq:Wstarsetcoveringg}. We do not pursue this, in the interest of space.}
\begin{remark}
Relative to maximum coverage problems, \cite{gairing2009covering} explicitly determines the price of anarchy for the mechanism in \eqref{eq:fstargair}. In the limit as $n\rightarrow\infty$ (i.e., when there is no bound on the number of agents), it's value amounts to 
\[
1-\frac{1}{e}
\]
and thus exactly matches the result in \eqref{eq:1-ce}, since for covering problems it is $c=1-(w(n)-w(n-1))=1$.
\end{remark}

\subsection{Application: content distribution in wireless data networks}
\label{subsec:content}
In this section we consider the problem of distributed data caching introduced in \cite{goemans2006market} as a technique to reduce peak traffic in mobile data networks. In order to alleviate the growing radio congestion caused by the recent surge of mobile data traffic, the latter work suggested to %
store popular and spectrum intensive items (such as movies or songs) in geographically distributed stations. 
The approach has the advantage of bringing the content closer to the customer, and to  avoid recurring transmission of large quantities of data.  
Similar offloading techniques, aiming at minimizing the peak traffic demand by storing popular items at local cells, have been recently proposed in the context of modern 5G mobile networks \cite{andrews2013seven}. The question we seek to answer in this section is how to  geographically distribute the popular items across the nodes of a network so as to maximize the total number of queries fulfilled.
In the following we borrow the model introduced in \cite{goemans2006market} and show how the utility design approach presented here yields improved theoretical and practical performances.

We consider a rectangular grid with $n_x\times n_y$ bins and a finite set $\mc{R}$ of data items. For each item $r\in\mc{R}$, we are given its query rate $q_r\ge0$ as well as its position in the grid $O_r$ and a radius $\rho_r$. A circle of radius $\rho_r$ centered in $O_r$ represents the region where the item $r$ is requested. Additionally we consider a set of geographically distributed nodes $N$ (the local cells), where each node $i\in N$ is assigned to a position in the grid $P_i$. A node is assigned a set of feasible allocations $\mc{A}_i$ according to the following rules:
\begin{itemize}
	\item[i)] $\mc{A}_i\subseteq 2^{\mc{R}_i}$, where $
	\mc{R}_i\coloneqq\{r\in\mc{R}~\text{s.t.}~||O_r- P_i||_2\le \rho_r\}$.
	That is, $r\in\mc{R}_i$ if the (euclidean) distance between the position of node $i$ and item $r$ is smaller equal to $\rho_r$.
	\item[ii)] $|\mc{A}_i|\le k_i$, for some $k_i\in \mb{N}$, $k_i\ge 1$.
\end{itemize}
In other words, node $i$ can include the resource $r$ in its allocation $a_i$ only if the node is in the region where the item $r$ is requested (first rule), while we limit the number of stored items to $k_i$ for reasons of space (second rule).\footnote{Similarly to what discussed for the application in Section \ref{sec:submod}, it is possible to reduce the problem to the case where $\mc{A}_i$ are the bases of a matroid $\mc{M}_i$. Since computing a single best response is a polynomial task (it amounts to sorting $q_r f(|a|_r)$ and picking the $k_i$ first items), Algorithm \ref{alg:BR} converges in polynomial time, owing to Proposition \ref{prop:poly} in Section~\ref{sec:complexity}.} The situation is exemplified in Figure \ref{fig:checkered}.
\begin{figure}[h!]
\begin{center}
\includegraphics[scale=0.8]{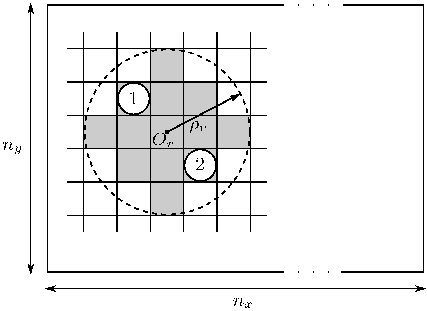}
\caption{The nodes $1$, $2$ can include the query $r$ in any allocation i.e. $r\in\mc{R}_1$ and $r\in\mc{R}_2$ since the distance from nodes $1$, $2$ to $O_r$ is less than~$\rho_r$.}
\label{fig:checkered}
\end{center}
\end{figure}

\noindent The objective is to select a feasible allocation for every node so as to jointly maximize the total amount of queries fulfilled,~i.e.,
\[
\max_{a\in\mc{A}}\sum_{r\in\cup_{i\in N} a_i}q_r\,.
\]
The work in  \cite{goemans2006market} proposes the use of the equal share mechanism $\fsv$ to obtain a distributed algorithm. 

In the following we compare the results of numerical simulations obtained using $\fsv$ or $\fopt$ in \eqref{eq:fstargair}. The following parameters are employed. %
We choose $n_x=n_y=800$, $|N|=100$, $|\mc{R}|=1000$. The nodes and the data items are uniformly randomly placed in the grid. The query rate of data items is chosen according to a power law (Zipf distribution) $q_r=1/r^\alpha$ for $r\in [1000]$.\footnote{Typical query rates follow this distribution, with $0.6\le \alpha\le 0.9$, see \cite{breslau1999web}.} The radii of interests are set to be identical for all items $\rho_r=\rho=200$. We let $k_i=5$ for all agents, while $\alpha$ varies in $0.6\le \alpha\le 0.9$. %
We consider $10^5$ instances of such problem, and for every instance compute a pure Nash equilibrium by means of Algorithm \ref{alg:BR}. Given the size of the problem, it is not possible to compute the optimal allocation and thus the price of anarchy. As a surrogate for the latter we use the ratio $W(\ae)/W_{\rm tot}$, where $\ae$ is the pure Nash equilibrium determined by the algorithm and
\[
W_{\rm tot}\coloneqq\sum_{r\in\mc{R}} q_r,\]
 is the sum of all the query rates, and thus is an upper bound for $W(\aopt)$. Observe that $W_{\rm tot}$ is a constant for all the simulations with fixed $\alpha$, indeed  $W_{\rm tot}= \sum_{r\le 1000}1/{r^\alpha}$ and thus serves as a mere scaling factor. The theoretical price of anarchy for large $n$ is $\poa(\fsv,\ones[n],n)=0.5$ (tight also when the query rates are Zipf distributed \cite{goemans2006market}) and $\poa(\fopt,\ones[n],n)=1-1/e \approx 0.632$.

Figure \ref{fig:Boxplot} compares the quantity $W(\ae)/W_{\rm tot}$ for the choice of $\fsv$ and $\fopt$, across different values of $\alpha$. First we observe that the worst cases encountered in the simulations are at least $10\%$ better than the theoretical counterparts. Further, for each fixed value of $\alpha$, there is a good separation between the performance of $\fsv$ and $\fopt$, in favor of the latter. This holds true, not only in the worst case sense (crosses in Figure \ref{fig:Boxplot}), but also on average. As $\alpha$ increases from $0.6$ to $0.9$, the worst case performance seems to degrade for both $\fsv$ and $\fopt$. Nevertheless, since we are using $W(\ae)/W_{\rm tot}$ as a surrogate for the true price of anarchy, it is unclear if the previous conclusion also holds for $W(\ae)/W(\aopt)$.

 \begin{figure}[ht!] 
\begin{center}
\hspace*{-5mm} 
\setlength\figureheight{4cm} 
\setlength\figurewidth{6.7cm} 
\input{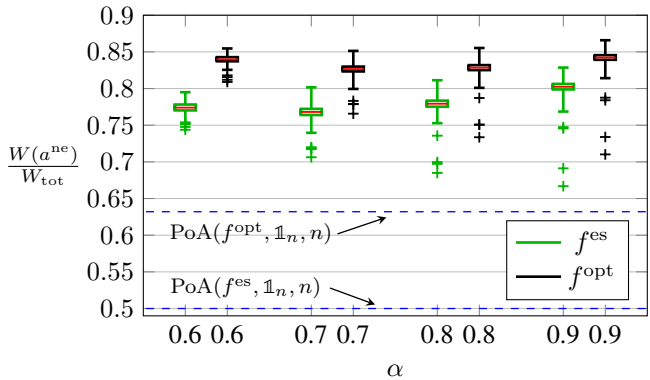}
\caption{Box plot comparing the performance of the best response algorithm on $10^5$ instances for the choice of mechanisms $\fsv$ and $\fopt$, across different values of $\alpha$. On each plot, the median is represented with a red line, and the corresponding box contains the 25th and 75th percentiles. The (four) worst cases are represented with crosses. \mbox{The dashed lines represent the price of anarchy.}} 
\label{fig:Boxplot}
\end{center}
\end{figure}
Figure \ref{fig:PDF_comparison_content} presents a more detailed comparison between $\fsv$ and $\fopt$ for a fixed value of $\alpha=0.7$ over all the $10^5$ instances. Relative to this case, Figure \ref{fig:BR_count} describes the (distribution of the) number of best response rounds required for Algorithm \ref{alg:BR} to converge. 
Quick convergence is achieved, with a number of best response rounds equal to $11$ in the worst case.
Observe that in every best response round all players have a chance to update their decision variable, so that a total number of $n_{\rm BR}$ rounds amounts to $n\cdotshort n_{\rm BR}$ individual best responses.
 
\begin{figure}[ht!] 
\begin{center}
\hspace*{-5mm} 
\setlength\figureheight{4cm} 
\setlength\figurewidth{6.7cm} 
\begin{tikzpicture}

\begin{axis}[%
width=\figurewidth,
height=\figureheight,,
at={(1.011in,0.642in)},
scale only axis,
xmin=0.49,
xmax=0.86,
xlabel={$W(\ae)/W_{\rm tot}$},
ymin=-300,
ymax=7500,
unbounded coords=jump,
scaled y ticks=base 10:-3,
%extra y tick style={1000},
%ytick={0, 2000, 4000, 6000, 8000},
%yticklabels={$0$, $2\cdotshort10^4$, $4\cdotshort10^4$, $6\cdotshort10^4$,
%	$8\cdotshort10^4$},
ylabel={Count},
ylabel style={rotate=-90},
grid=both,
legend style={at={(0.044,0.165)}, anchor=west, draw=white!15!black},
legend cell align={left}
]
%% vertical lines
\addplot[color=blue, dashed , line width = 0.5pt, forget plot]
table[row sep=crcr]
  {0.5 	-2000\\
   0.5 	10000\\};
\addplot[color=blue, dashed , line width = 0.5pt, forget plot]
table[row sep=crcr]
  {0.632 	-2000\\
   0.632	10000\\};

%   \addplot[color=blue, dashed , line width = 1pt, forget plot]
%table[row sep=crcr]
%  {0.556 	-0.1\\
%   0.556 	1\\};
%   \addplot[color=black, dashed , line width = 1pt, forget plot]
%table[row sep=crcr]
%  {0.688 	-0.1\\
%   0.688 	1\\};
%% Plots of single point to get desired legend
%%
\addplot[color=green, mark = triangle, mark size = 4pt, mark options={solid}, line width = 1pt]
  table[row sep=crcr]
  {0.706274043136323	0\\};
\addlegendentry{$\fsv$\hspace*{-1mm}};
%
%\addplot [color=blue, mark = x, mark size = 4pt, line width = 1pt]
%  table[row sep=crcr]
%  {%
%0.71459	2.5e-05\\};
%\addlegendentry{$\fmc$\hspace*{-1mm}};
%
\addplot[color=black, mark = x, mark size = 4pt, line width = 1pt]
  table[row sep=crcr]{%
0.765721086795839	0\\};
\addlegendentry{$\fopt$};

\addplot[ybar interval, fill=green, fill opacity=0.6, draw=none, area legend, line width=0.05pt] table[row sep=crcr] {%
x	y\\
0.7056	1\\
0.706561	0\\
0.707522	0\\
0.708483	0\\
0.709444	0\\
0.710405	0\\
0.711366	0\\
0.712327	0\\
0.713288	0\\
0.714249	0\\
0.71521	0\\
0.716171	0\\
0.717132	1\\
0.718093	1\\
0.719054	1\\
0.720015	0\\
0.720976	0\\
0.721937	0\\
0.722898	0\\
0.723859	0\\
0.72482	0\\
0.725781	0\\
0.726742	0\\
0.727703	0\\
0.728664	0\\
0.729625	0\\
0.730586	0\\
0.731547	0\\
0.732508	0\\
0.733469	0\\
0.73443	0\\
0.735391	0\\
0.736352	0\\
0.737313	0\\
0.738274	0\\
0.739235	1\\
0.740196	0\\
0.741157	2\\
0.742118	0\\
0.743079	2\\
0.74404	1\\
0.745001	2\\
0.745962	5\\
0.746923	4\\
0.747884	18\\
0.748845	25\\
0.749806	49\\
0.750767	65\\
0.751728	117\\
0.752689	244\\
0.75365	439\\
0.754611	576\\
0.755572	632\\
0.756533	1081\\
0.757494	1588\\
0.758455	2313\\
0.759416	2802\\
0.760377	2760\\
0.761338	3551\\
0.762299	4660\\
0.76326	5513\\
0.764221	6133\\
0.765182	5555\\
0.766143	5651\\
0.767104	6105\\
0.768065	6289\\
0.769026	6288\\
0.769987	5789\\
0.770948	4872\\
0.771909	4449\\
0.77287	3998\\
0.773831	3595\\
0.774792	3164\\
0.775753	2585\\
0.776714	2019\\
0.777675	1593\\
0.778636	1355\\
0.779597	1052\\
0.780558	818\\
0.781519	591\\
0.78248	450\\
0.783441	323\\
0.784402	248\\
0.785363	194\\
0.786324	147\\
0.787285	92\\
0.788246	63\\
0.789207	42\\
0.790168	35\\
0.791129	13\\
0.79209	11\\
0.793051	9\\
0.794012	6\\
0.794973	1\\
0.795934	1\\
0.796895	3\\
0.797856	3\\
0.798817	1\\
0.799778	2\\
0.800739	1\\
0.8017	1\\
};
\addplot[ybar interval, fill=black, fill opacity=0.6, line width=0.05pt, draw=none, area legend] table[row sep=crcr] {%
x	y\\
0.7656	1\\
0.766458	0\\
0.767316	0\\
0.768174	0\\
0.769032	0\\
0.76989	0\\
0.770748	0\\
0.771606	0\\
0.772464	0\\
0.773322	0\\
0.77418	0\\
0.775038	0\\
0.775896	0\\
0.776754	0\\
0.777612	0\\
0.77847	2\\
0.779328	0\\
0.780186	0\\
0.781044	0\\
0.781902	0\\
0.78276	1\\
0.783618	0\\
0.784476	0\\
0.785334	0\\
0.786192	0\\
0.78705	0\\
0.787908	0\\
0.788766	0\\
0.789624	0\\
0.790482	0\\
0.79134	0\\
0.792198	0\\
0.793056	0\\
0.793914	0\\
0.794772	0\\
0.79563	0\\
0.796488	1\\
0.797346	0\\
0.798204	0\\
0.799062	1\\
0.79992	1\\
0.800778	0\\
0.801636	0\\
0.802494	0\\
0.803352	1\\
0.80421	1\\
0.805068	2\\
0.805926	0\\
0.806784	3\\
0.807642	1\\
0.8085	5\\
0.809358	8\\
0.810216	8\\
0.811074	17\\
0.811932	40\\
0.81279	88\\
0.813648	129\\
0.814506	249\\
0.815364	506\\
0.816222	815\\
0.81708	1342\\
0.817938	1820\\
0.818796	2119\\
0.819654	2837\\
0.820512	3648\\
0.82137	4672\\
0.822228	5632\\
0.823086	6254\\
0.823944	6115\\
0.824802	6473\\
0.82566	6677\\
0.826518	6932\\
0.827376	6906\\
0.828234	6255\\
0.829092	5598\\
0.82995	4796\\
0.830808	3974\\
0.831666	3503\\
0.832524	3019\\
0.833382	2413\\
0.83424	1929\\
0.835098	1439\\
0.835956	1062\\
0.836814	752\\
0.837672	593\\
0.83853	431\\
0.839388	290\\
0.840246	206\\
0.841104	149\\
0.841962	101\\
0.84282	66\\
0.843678	36\\
0.844536	38\\
0.845394	17\\
0.846252	9\\
0.84711	3\\
0.847968	6\\
0.848826	2\\
0.849684	1\\
0.850542	5\\
0.8514	5\\
};
\end{axis}

%text and arrows
\draw[-stealth, line width=0.5pt] (5.55 ,5.15) -- (5.25, 4.4);
\draw[-stealth, line width=0.5pt] (3.15 ,5.15) -- (2.85, 4.4);
\node at (6.45, 5.3) {\footnotesize $\poa(\fopt,\ones[n],n)$};
\node at (4, 5.3) {\footnotesize $\poa(\fsv,\ones[n],n)$};
%%
%\node at (4, 2.5) {\footnotesize $\poa(\fmc)$};
%\draw[-stealth, line width=0.5pt] (3.25 ,2.5) -- (2.85, 2.5);
%\node at (4, 3) {\footnotesize $\poa(\fsv)$};
%\draw[-stealth, line width=0.5pt] (3.25 ,3) -- (3.05, 3);
\end{tikzpicture}% 
\caption{Distribution of $W(\ae)/W_{\rm tot}$ on $10^5$ instances for fixed $\alpha=0.7$.
The dashed lines represent the value of $\poa(\fsv,w,n)$, $\poa(\fopt,w,n)$, while the corresponding markers identify the worst case performance encountered during the simulations.
} %
\label{fig:PDF_comparison_content}
\end{center}
\end{figure}
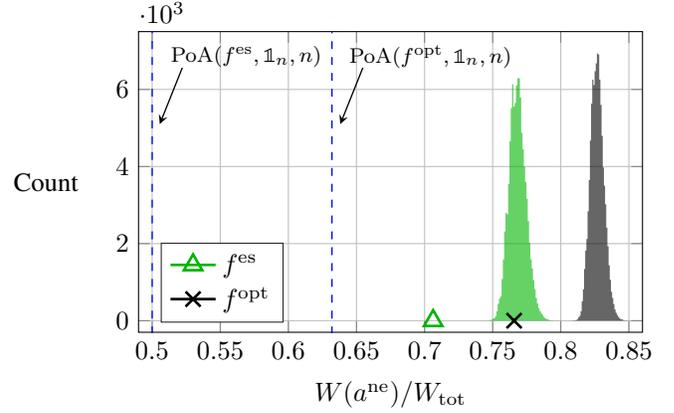
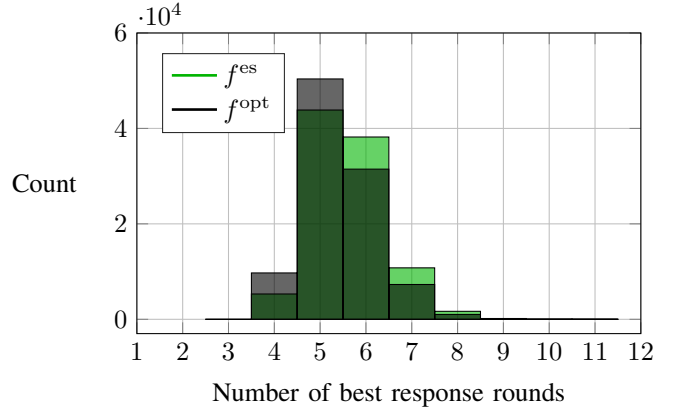
\begin{figure}[ht!] 
\begin{center}
\hspace*{-5mm} 
\setlength\figureheight{4cm} 
\setlength\figurewidth{6.7cm} 
\begin{tikzpicture}

\begin{axis}[%
width=\figurewidth,
height=\figureheight,,
at={(1.011in,0.642in)},
scale only axis,
xmin=1,
xmax=12,
xlabel={Number of best response rounds},
ymin=-3000,
ymax=60000,
unbounded coords=jump,
%ytick={0, 2000, 4000, 6000, 8000},
%%yticklabels={$0$, $2\cdotshort10^4$, $4\cdotshort10^4$, $6\cdotshort10^4$,
%	$8\cdotshort10^4$},
xtick={1, 2, 3, 4, 5, 6, 7, 8, 9, 10, 11, 12},
ylabel={Count},
ylabel style={rotate=-90},
grid=both,
legend style={at={(0.05,0.8)}, anchor=west, draw=white!15!black},
legend cell align={left}
]
%% vertical lines
%\addplot[color=green, dashed , line width = 1pt, forget plot]
%table[row sep=crcr]
%  {0.568 	-0.1\\
%   0.568 	1\\};
%   \addplot[color=blue, dashed , line width = 1pt, forget plot]
%table[row sep=crcr]
%  {0.556 	-0.1\\
%   0.556 	1\\};
%   \addplot[color=black, dashed , line width = 1pt, forget plot]
%table[row sep=crcr]
%  {0.688 	-0.1\\
%   0.688 	1\\};
%% Plots of single point to get desired legend
%%
\addplot[color=green, line width = 1pt]
  table[row sep=crcr]
  {0.706274043136323	0\\};
\addlegendentry{$\fsv$\hspace*{-1mm}};
%
%\addplot [color=blue, mark = x, mark size = 4pt, line width = 1pt]
%  table[row sep=crcr]
%  {%
%0.71459	2.5e-05\\};
%\addlegendentry{$\fmc$\hspace*{-1mm}};
%
\addplot[color=black, line width = 1pt]
  table[row sep=crcr]{%
0.765721086795839	0\\};
\addlegendentry{$\fopt$};

\addplot[ybar interval, fill=green, fill opacity=0.6, draw=black, area legend] table[row sep=crcr] {%
x	y\\
3.5	5304\\
4.5	43853\\
5.5	38200\\
6.5	10781\\
7.5	1673\\
8.5	168\\
9.5	19\\
10.5	2\\
11.5	2\\
};
\addplot[ybar interval, fill=black, fill opacity=0.6, draw=black, area legend] table[row sep=crcr] {%
x	y\\
2.5	9\\
3.5	9715\\
4.5	50367\\
5.5	31468\\
6.5	7295\\
7.5	1018\\
8.5	119\\
9.5	8\\
10.5	1\\
11.5	1\\
};
\end{axis}
%text and arrows
%\draw[-stealth, line width=0.5pt] (5.2 ,2.5) -- (4.8, 2.5);
%\node at (5.9, 2.5) {\footnotesize $\poa(f^\star)$};
%%
%\node at (4, 2.5) {\footnotesize $\poa(\fmc)$};
%\draw[-stealth, line width=0.5pt] (3.25 ,2.5) -- (2.85, 2.5);
%\node at (4, 3) {\footnotesize $\poa(\fsv)$};
%\draw[-stealth, line width=0.5pt] (3.25 ,3) -- (3.05, 3);
\end{tikzpicture}% 
\caption{Distribution of the number of best response rounds required for convergence on $10^5$ instances, $\alpha=0.7$. The average number of best response rounds is $5.603$ and $5.39$ for the equal share and optimal mechanism, respectively.} 
\label{fig:BR_count}
\end{center}
\end{figure}

\section{The case of supermodular welfare function}
\label{sec:supermod}
In this section we consider welfare basis functions that are non-decreasing and convex, resulting in a monotone and supermodular total welfare $W(a)$. Applications featuring this property include but are not limited to clustering and image segmentation \cite{stobbe2010efficient}, power allocation in multiuser networks \cite{yassin2017centralized}. 
In the following we explicitly characterize the price of anarchy for the class of supermodular resource allocation problems as a function of $f$ (Theorem \ref{thm:convex}). Additionally, we show that the equal share mechanism maximizes this measure of efficiency, but \emph{is not the only one}. Finally, we show how our approach recovers and generalizes the results in \cite{phillips2017design, jensen2018}.

\begin{assumption}
\label{ass:submodd}
Assume that $w\in\mb{R}^n_{>0}$ is a non-decreasing and convex function, that is %
\[
\begin{split}
w(j+1)&\ge w(j)\,,\\%\quad \forall j\in [n-1]\\
w(j+1)-w(j)&\ge w(j)-w(j-1)\,,\quad \forall j\in [n-1]\,,
\end{split}
\]	
where we define $w(0)=0$.
\end{assumption}

\begin{theorem}[\bf $ \rm{\bf PoA}$ for supermodular welfare]
\label{thm:convex}
Let Assumption \ref{ass:fw=1} hold.
Further, let $w$ satisfy Assumption \ref{ass:submodd}.
Consider a mechanism $f\in\mb{R}^n$ such that $f(j)\ge 1$ for all $j \in[n]$. It holds 
\[
\poa{(f,w,n)}= \frac{n}{w(n)} \frac{1}{\max_{j\in[n]} j \frac{f(j)}{w(j)}}\,.
\]
It follows that $\fsv$ is optimal and achieves
\[
\poa{({\fsv},w,n)}= \frac{n}{w(n)}.
\]
\end{theorem}
Observe that the equal share mechanism satisfies the conditions of Theorem \ref{thm:convex}, since for all $j\in[n]$ it is $\fsv(j)=w(j)/j\ge 1$, where $w(j)/j \ge 1$ by convexity of $w$ and by $w(1)=1$.
Further note that the equal share mechanism is \emph{not} the unique maximizer of $\poa(f,w,n)$. Indeed, all the mechanisms with $f(1)=1$ and $1\le f(j)\le w(j)/j$, $j\in[n]$, are optimal, since the previous theorem applies and they have $\max_{j\in[n]} j{f(j)}/{w(j)}=1$ due to $f(1)=w(1)=1$ and $jf(j)/w(j)\le 1$ for $j\in[n]$.
\begin{remark}
Relative to concave cost-sharing games and convex welfare-maximization problems,  \cite{phillips2017design} and \cite{jensen2018} include partial expressions for the price of anarchy of specific mechanisms (e.g., the equal share). Additionally, \cite{phillips2017design,jensen2018} show that $\fsv$ maximizes such efficiency metric. The expression of $\poa(f,w,n)$ obtained in Theorem \ref{thm:convex} generalizes the result of \cite{jensen2018}, and that of \cite{phillips2017design} to any mechanism.
\end{remark}
Figure \ref{fig:comparisonsuperoriginal} compares the price of anarchy of the Shapley value, marginal contribution and optimal mechanisms (derived as the solution of \eqref{eq:proppart2}), in the case when $w(j)=j^d$ with $1\le d\le 2$, $|N|\le 20$. First, we observe that the solution of \eqref{eq:proppart2} and $\fsv$ give the same performance, as predicted from Theorem \ref{thm:convex}. Additionally, we observe that the quality of the approximation quickly degrades as the welfare basis $w$ gets steeper ($d$ gets larger). This is due to the fact that if $w(n)$ grows much faster than $n$, the quantity $n/w(n)$ quickly decreases.
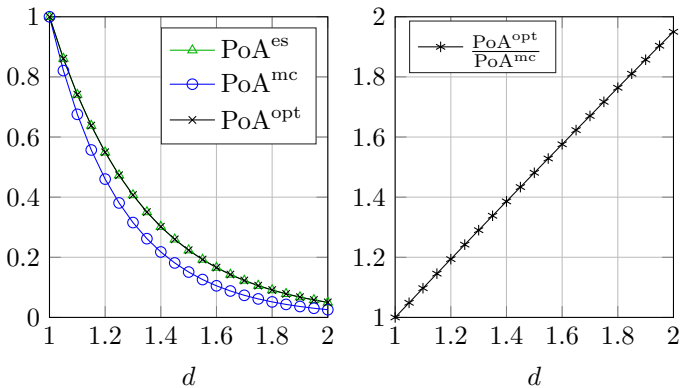
\begin{figure}[ht!] 
\centering
\hspace*{-5mm}
\setlength\figureheight{4cm} 
\setlength\figurewidth{3.7cm} 
\begin{tikzpicture}
\begin{axis}[%
width=\figurewidth,
height=\figureheight,,
at={(1.011in,0.642in)},
scale only axis,
xmin=1,
xmax=2,
xlabel={$d$},
ymin=0,
ymax=1,
grid=both,
legend style={at={(0.4,0.77)}, anchor=west, draw=white!15!black},
legend cell align={left}
]

\addplot[color=green,mark=triangle,mark options={solid}]
  table[row sep=crcr]{%
1	1\\
1.05	0.860891659331735\\
1.1	0.741134449106947\\
1.15	0.638036465679592\\
1.2	0.549280271653059\\
1.25	0.472870804501588\\
1.3	0.407090531536904\\
1.35	0.350460843193043\\
1.4	0.301708816827258\\
1.45	0.259738603953433\\
1.5	0.223606797749979\\
1.55	0.192501227152835\\
1.6	0.165722700866999\\
1.65	0.142669290938328\\
1.7	0.122822802611579\\
1.75	0.105737126344056\\
1.8	0.091028210151304\\
1.85	0.078365426883154\\
1.9	0.0674641423836782\\
1.95	0.0580793174820771\\
2	0.05\\
};
\addlegendentry{$\poa^{\rm es}$};
\addplot [color=blue,mark=o,mark options={solid}]
  table[row sep=crcr]{%
1	1\\
1.05	0.820939648857225\\
1.1	0.675473223027811\\
1.15	0.556933108559613\\
1.2	0.46006526711858\\
1.25	0.380706470829292\\
1.3	0.315541335661298\\
1.35	0.261917758663104\\
1.4	0.217705571084728\\
1.45	0.181187487420064\\
1.5	0.150974407561143\\
1.55	0.125939233377645\\
1.6	0.105164866104713\\
1.65	0.0879031394436211\\
1.7	0.0735422385807872\\
1.75	0.0615807418696769\\
1.8	0.0516068582576294\\
1.85	0.043281760749507\\
1.9	0.0363261634019941\\
1.95	0.0305094773888512\\
2	0.0256410256410256\\};
\addlegendentry{$\poa^{\rm mc}$};
%\addplot [color=red,mark=square,mark options={solid}]
%  table[row sep=crcr]{%
%%
%0	0.632120558828558\\
%0.05	0.63290745689283\\
%0.1	0.633948500190576\\
%0.15	0.635306399990088\\
%0.2	0.637058289861445\\
%0.25	0.639299182932335\\
%0.3	0.642146390943544\\
%0.35	0.645745237489006\\
%0.4	0.650276538858866\\
%0.45	0.655966537117448\\
%0.5	0.663100291363061\\
%0.55	0.672040030296724\\
%0.6	0.683250752894955\\
%0.65	0.697336624484692\\
%0.7	0.715093788519244\\
%0.75	0.73758870772346\\
%0.8	0.766277195099631\\
%0.85	0.803190075166655\\
%0.9	0.851231268101469\\
%0.95	0.914671985458154\\
%0.9999	0.999801547920889\\
%};
%\addlegendentry{$\fmc$};

\addplot[color=black,solid,mark=x,mark options={solid}]
  table[row sep=crcr]{%
1	1\\
1.05	0.860891659331735\\
1.1	0.741134449106947\\
1.15	0.638036465679592\\
1.2	0.549280271653059\\
1.25	0.472870804501588\\
1.3	0.407090531536904\\
1.35	0.350460843193043\\
1.4	0.301708816827258\\
1.45	0.259738603953433\\
1.5	0.223606797749979\\
1.55	0.192501227152835\\
1.6	0.165722700866999\\
1.65	0.142669290938328\\
1.7	0.122822802611579\\
1.75	0.105737126344056\\
1.8	0.091028210151304\\
1.85	0.078365426883154\\
1.9	0.0674641423836782\\
1.95	0.0580793174820771\\
2	0.05\\
};
\addlegendentry{$\poa^{\rm opt}$};
\end{axis}
\end{tikzpicture}%
\begin{tikzpicture}
\begin{axis}[%
width=\figurewidth,
height=\figureheight,,
at={(1.011in,0.642in)},
scale only axis,
xmin=1,
xmax=2,
xlabel={$d$},
ymin=1,
ymax=2,
grid=both,
legend style={at={(0.05,0.9)}, anchor=west, draw=white!15!black},
legend cell align={left}
]
\addplot [color=black,mark=asterisk,mark options={solid}]
  table[row sep=crcr]{%
1	1\\
1.05	1.048666197\\
1.1	1.097207741\\
1.15	1.145624952\\
1.2	1.193918148\\
1.25	1.242087647\\
1.3	1.290133765\\
1.35	1.33805682\\
1.4	1.385857125\\
1.45	1.433534995\\
1.5	1.481090745\\
1.55	1.528524686\\
1.6	1.575837131\\
1.65	1.623028391\\
1.7	1.670098776\\
1.75	1.717048596\\
1.8	1.763878159\\
1.85	1.810587775\\
1.9	1.857177749\\
1.95	1.903648389\\
2	1.95\\};
\addlegendentry{$\frac{\poa^{\rm opt}}{\poa^{\rm mc}}$};
%\addplot[color=black,solid,mark=x,mark options={solid}]
%  table[row sep=crcr]{%
%1	1\\
%1.05	0.860891659331735\\
%1.1	0.741134449106947\\
%1.15	0.638036465679592\\
%1.2	0.549280271653059\\
%1.25	0.472870804501588\\
%1.3	0.407090531536904\\
%1.35	0.350460843193043\\
%1.4	0.301708816827258\\
%1.45	0.259738603953433\\
%1.5	0.223606797749979\\
%1.55	0.192501227152835\\
%1.6	0.165722700866999\\
%1.65	0.142669290938328\\
%1.7	0.122822802611579\\
%1.75	0.105737126344056\\
%1.8	0.091028210151304\\
%1.85	0.078365426883154\\
%1.9	0.0674641423836782\\
%1.95	0.0580793174820771\\
%2	0.05\\
%};
%\addlegendentry{$\poa^{\rm opt}$};
\end{axis}
\end{tikzpicture}% 
\caption{Left: comparison between the price of anarchy of the optimal mechanism $\fopt$ (determined as the solution of \eqref{eq:proppart2}),
equal share $\fsv$, and marginal contribution $\fmc$ mechanisms, denoted with  $\poa^{\rm opt}$, $\poa^{\rm es}$ and $\poa^{\rm mc}$, respectively. Note how $\fopt$ and $\fsv$ have identical price of anarchy, as predicted by Theorem \ref{thm:convex}.
Right: relative performance of the optimal mechanism compared to the marginal contribution.
The problems considered features $|N|\le 20$ and a welfare basis of the form $w(j)=j^d$ with $d\in[1,2]$ represented over the $x$-axis.
} 
\label{fig:comparisonsuperoriginal}
\end{figure}

\section{Complexity of computing pure Nash equilibria}
\label{sec:complexity}

Since the price of anarchy represents the \emph{approximation ratio} of any algorithm capable of computing a pure Nash equilibrium, it is important to understand whether it is possible to compute one such equilibrium efficiently.
Even though many equilibrium-computing algorithms have been proposed, in the following we focus on the \emph{round-robin best response dynamics} (Algorithm \ref{alg:BR}), presented in Section \ref{sec:problemformulation}.

While computing a pure Nash equilibrium is a $\mc{NP}$-hard task for a general game \cite{gottlob2005pure}, all the instances $G$ considered in this work are congestion games as observed in Appendix \ref{app:congestion}. Relative to this class of games, the following proposition provides sufficient conditions under which Algorithm \ref{alg:BR} has polynomial running time. The main assumption amounts to requiring that the agents' allocation sets coincide with the set of bases of a matroid. 
The notion of matroid generalizes that of linear independence for vector spaces.
The definition of matroid, its rank, \mbox{and related notions can be found in Appendix~\ref{appendixa}.}
\begin{proposition}
\label{prop:poly}\cite[Thm. 2.5]{ackermann2008impact}
Consider the congestion game $G$ and assume the allocation sets $\mc{A}_1,\mc{A}_2,\dots,\mc{A}_n$ are the set of bases for some matroid over the set of resources $\mc{M}_i=(\mc{R},\mc{I}_i)$, where $\mc{I}_i\subseteq 2^\mc{R}$, $|\mc{R}|=m$. Then, the best response dynamics (Algorithm \ref{alg:BR}) reaches a pure Nash equilibrium after at most $n^2 m\,\max_{i\in N}\text{rank}(\mc{M}_i)$ best response rounds.	
\end{proposition}

\noindent {\bf Examples of Matroids.}
The case when each feasible allocation consists of a single resource does satisfy the assumptions of the previous theorem, even if an agent does not have access to all the possible resources. One such example is the following: $\mc{R}=\{r_1,\dots,r_n, r_{n+1}\}$, $\mc{A}_i=\{\{r_i\},\{r_{n+1}\}\}$. Define $\mc{I}_i=\{\emptyset,\{r_i\},\{r_{n+1}\}\}$. We have that $\mc{M}_i\coloneqq(\mc{R},\mc{I}_i)$ is a matroid of rank $1$ and that $\mc{A}_i$ is a set of bases for $\mc{M}_i$, see Appendix \ref{appendixa} for the details.
In the same Appendix, we provide an example of allocation sets satisfying the property required in the previous proposition (the case of uniform matroid).
On the negative side, it is simple to construct examples that \emph{do not} satisfy this requirement. For instance, consider $\mc{R}=\{r_1,\dots,r_m\}$, $m\ge3$ and $\mc{A}_1=\{\{r_1\},\{r_2,r_3\}\}$. The set $\mc{A}_1$ cannot be the set of bases for any matroid $\mc{M}_1$, as all bases must have the same number of elements (see Appendix \ref{appendixa}) while $\{r_1\}$ and $\{r_2,r_3\}$ do not. %
\begin{remark}
 The previous proposition gives condition under which the maximum number of best responses required to converge to a pure Nash equilibrium is polynomially bounded in the number of players and resources. If it is possible to compute a single best response polynomially in the number of resources, then the performance guarantee given by $\poa$ is achievable in \emph{polynomial time} by Algorithm \ref{alg:BR}.	The applications presented in Sections \ref{sec:submod}, \ref{sec:coveringg} satisfy these assumptions.
\end{remark}

\section{Conclusions and Remarks}
This manuscript specializes the results presented in Part I to the case of monotone submodular, supermodular and covering problems. 
For each class of problems, we derived an explicit characterization of the price of anarchy as a function of the assigned utility functions.
We then compared the performance provided by optimally designed mechanisms with existing approximation results. For covering problems we recovered the $1-1/e$ result of \cite{nemhauser1978analysis}, while in the submodular case our bounds improve on the recent approximation result of \cite{sviridenko2017optimal}. 
Finally, we tested the theoretical findings on two applications.

We remark on the fact that the performance certificates obtained in this work are confined to the notion of pure Nash equilibrium. While computing one such equilibrium is, in general, a hard task, Proposition \ref{prop:poly} showed that under structural assumptions on the sets $\{\mc{A}_i\}_{i\in N}$, this can be accomplished in polynomial time.
Whether the performance certificates derived here hold for the larger class of coarse correlated equilibria (CCE), is at this point an open question. In this respect, Remark \ref{rmk:comparisontim} shows that this is the case limitedly to $\fsv$. The appeal of CCE lies in the fact that their calculation is a polynomial task for the classes of games considered here~\cite{papadimitriou2008computing}. 

\appendices
\appendices
\section{Congestion games}
\label{app:congestion}
In this section we recall the definition of congestion games, and corresponding properties. While these are typically defined for cost-minimization problems, we provide here the definition of congestion games relative to welfare-maximization games.

\begin{definition}[Congestion game, \cite{rosenthal1973class}]
\label{def:congestiongame}
Consider $\mc{R}$ a finite set of resources and for every resource $r\in\mc{R}$ a function $f_r:\mb{N}\rightarrow\mb{R}$. A congestion game is game where $\N=\{1,\dots,n\}$ is the set of players, $\mc{A}_i\subseteq 2^{\mc{R}}$ and $U_i(a)=\sum_{r\in a_i} f_r(|a|_r)$ are the action set and utility function of player $i$, respectively. The expression $|a|_r$ denotes the number of players selecting resource $r$ in allocation $a$, $|a|_r\coloneqq\{i\in\N\,\text{s.t.}\,r\in a_i\}$.
\end{definition}

\noindent It is immediate to observe that any game $G$ defined in \eqref{eq:gameG} is a congestion game, regardless of what mechanism $f$ is chosen.

 Owing to their structure, congestion games belongs to the class of potential games \cite{monderer1996potential}. As a consequence, a pure Nash equilibrium is guaranteed to exist \cite{rosenthal1973class}. Additionally, the best response dynamics described in Algorithm \ref{alg:BR} always converges to  a pure Nash equilibrium in a finite number of rounds \cite{monderer1996potential}.

\section{Matroids}
The notion of matroid was introduced by Whitney \cite{Whitney35} with the objective of generalizing the notion of linear independence from vector spaces to more abstract structures. 
Since then, matroids have found a number of applications, most notably to combinatorial optimization, graph and network theory.
Informally, a matroid is a collections of subsets $\mc{I}$ of a finite set $\mc{R}$ that satisfies two properties: (i) any subset $A$ of a given set $B\in\mc{I}$ also belongs to $\mc{I}$ (hereditary property); (ii) if two sets $A\in\mc{I}$ and $B\in\mc{I}$ have different cardinality $|B|>|A|$, there must exist an element $\{r\}$ belonging to their difference, such that $A$ augmented with $\{r\}$ is a set in $\mc{I}$ (augmentation property).
Their formal definition follows.
\begin{definition}[Matroid]
\label{def:matroid}
A tuple $\mc{M}=(\mc{R},\mc{I})$ is a matroid if $\mc{R}$ is a finite set, $\mc{I}\subseteq 2^\mc{R}$ is a collection of subsets of $\mc{R}$, and the following two properties hold:
\begin{itemize}
	\item If $B\in\mc{I}$ and $A\subseteq B$, then $A\in\mc{I}$;
	\item If $A\in\mc{I}$, $B\in\mc{I}$ and $|B|>|A|$, then there exists an element $r\in B\setminus A$ s.t. $A\cup \{r\}\in\mc{I}$.
\end{itemize}	
\end{definition}
\begin{definition}[Basis]
	Given a matroid $\mc{M}=(\mc{R},\mc{I})$, a set $S\in\mc{I}$ such that for all $r\in\mc{R}\setminus S$, $(S\cup\{r\})\notin \mc{I}$ is called a basis of the matroid $\mc{M}$. 
\end{definition}
\noindent
It can be shown that all basis have the same number of elements, which is known as the rank of the matroid, and indicated with $\text{rank}(\mc{M})$, see \cite{welsh2010matroid}.

A well-known example of matroid is that of a finite subset of a vector space (corresponding to the set $\mc{R}$), together with any set of linearly independent vectors belonging to $\mc{R}$ (corresponding to the collection $\mc{I}$).
Another common example of matroid is that of uniform matroid defined as follows.
\begin{definition}[Uniform matroid]
	Given a finite set $\mc{R}$ with $|\mc{R}|=m$, let $\mc{I}\subseteq 2^\mc{R}$ be the collection of all subsets with a number of elements $k\le m$. $\mc{M}=(\mc{R},\mc{I})$ is a matroid and it is called the uniform matroid of rank$(\mc{M})=k$, see \cite{welsh2010matroid}.
\end{definition}

\label{appendixa}
\section{Proof of Theorem \ref{thm:wconcavefwdecreas}}
\label{appendixb}
\begin{proof}
If $f(1)\le0$, then $\poa(f,w,n)=0$ by Theorem \ref{thm:mine}. Thus, in the following we focus on the case of $f(1)>0$. Observe that the claim we wish to prove (i.e., the value of $W^\star$ in \eqref{eq:poasubmodular}) can be equivalently reformulated as in the following program, upon observing that for $j+l\le n$ it holds $\min(j,n-l)=j$ and $\min(l,n-j)=l$, while for $j+l> n$ it holds $\min(j,n-l)=n-l$ and $\min(l,n-j)=n-j$,
\be
\label{eq:programreducedproof}
\small
	\begin{split}
	W^\star &= \min_{\mu\in\mb{R}}~ \mu  \\[0.1cm]
	&\,\text{s.t.}~ \,\mu w(j)\!\ge \!w(l)+j f(j)\!-\!l f(j\!+\!1)\\%1\le j+l\le k\\
	&~\hspace*{25mm} \forall j,l\in[n]~{s.t.}~ j\ge l ~~\text{and}~~j+l\le n,\\[0.15cm]
	&\!\qquad \mu w(j)\!\ge\! w(l)\!+\!(n\!-\!l) f(j)\!-\!(n\!-\!j) f(j\!+\!1) \\
	&~\hspace*{25mm} \forall j,l\in[n]~{s.t.}~ j\ge l ~~\text{and}~~ j+l> n\,.
	\end{split}
\ee
In the following we prove that the latter program follows from \cite[Eq. \eqref{eq:formulacor1}]{part1Paccagnan2018} by showing that only the constraints with $l \le j$ and $l\ge 1$ are required, and that the decision variable $\lambda$ appearing in \cite[Eq. \eqref{eq:formulacor1}]{part1Paccagnan2018} takes the value $\lambda^\star=1$.

First, notice that $f(j)$ is assumed to be non-increasing, so that $W^\star$ can be correctly computed using \cite[Eq. \eqref{eq:formulacor1}]{part1Paccagnan2018}.	
For $j=0$, the constraints in \cite[Eq. \eqref{eq:formulacor1}]{part1Paccagnan2018}
 read as 
\[\lambda\ge\frac{w(l)}{l}\quad\forall \,l\in [n]\,,
\]	
(due to $f(1)=1$), and the most binding amounts to $\lambda\ge 1$ (due the to concavity of $w$ and to $w(1)=1$). 
For $j\neq 0$, we intend to show that the constraints with $l>j$ appearing in \cite[Eq. \eqref{eq:formulacor1}]{part1Paccagnan2018} are not required since
those with $j=l$ are more binding. 
Figure \ref{fig:explanation} illustrates this.
\begin{figure}[h!]
\centering
	\includegraphics[scale=1]{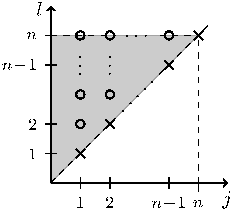}	\caption{Part {\bf i)} and {\bf ii)} of the proof amount to showing that for any constraint identified with the indices $(j,l)$ and $l>j>0$ (circles in the figure), the constraint identified with $(j,j)$ is more binding (crosses in the figure).}
\label{fig:explanation}
\end{figure}

To do so, we divide the discussion in the following two cases: {\bf i)} $1\le l+j\le n$ and  {\bf ii)} $l+j>n$.
\setlist[enumerate]{leftmargin=*}
\begin{enumerate}
\item[\bf i)]
For $1\le j+l\le n$ and $l>j$, we want to show that for any $\lambda\ge 1$ %
\[
w(j)+\lambda {j}[f(j)-f(j+1)]
\ge
{w(l)}+\lambda \left[jf(j)-lf(j+1)\right],
\]
where the left hand side is obtained setting $l=j$.
This is equivalent to showing
\begin{equation}
\label{eq:first}
w(l)-w(j)+\lambda (j-l)f(j+1)\le 0\,.
\end{equation}
By concavity of $w$ and  $l>j$, one observes that 
\[\begin{split}
w(l)&\le w(j+1)+(w(j+1)-w(j))(l-j-1)\\
&=w(j)+(w(j+1)-w(j))(l-j)
\end{split}
\]
and since $l-j>0$, $w(j+1)-w(j)\ge 0$, $\lambda\ge 1$, it holds
\be
w(l)\le w(j)+\lambda (w(j+1)-w(j))(l-j).
\label{eq:ineqproofT4}
\ee
Using inequality \eqref{eq:ineqproofT4}, the inequality in \eqref{eq:first} follows, since
\[
\begin{split}
&w(l)-w(j)+\lambda (j-l)f(j+1)\le\\
&\le w(j)+\lambda (w(j+1)-w(j))(l-j)\\
&\qquad\quad~-w(j)+\lambda (j-l)f(j+1)\
\\
&=\lambda(l-j)(w(j+1)-w(j)-f(j+1))\le 0\,,	
\end{split}
\]
where the last inequality holds because $f(j+1)\ge w(j+1)-w(j)$ (by assumption) and $l>j$. Observe that the previous inequality is never evaluated for $j=n$, as there is no $l\in[n]$ with $l>j=n$.
\item[\bf ii)] 
We now consider the case $j+l>n$ and $l>j$. We first consider the case of $j\ge n/2$. Here we intend to prove that  for any $\lambda \ge 1$
\[
\begin{split}
&w(j)+\lambda (n-j)[f(j)-f(j+1)]
\ge\\
&w(l)+\lambda [(n-l)f(j)-(n-j)f(j+1)],
\end{split}
\]
where the left hand side is obtained setting $l=j$.
The latter is equivalent to
\be
\label{eq:provedfirst}
w(l)-w(j)+\lambda(j-l)f(j)\le 0.
\ee
Similarly to \eqref{eq:ineqproofT4}, one can show that
\[
w(l)\le w(j)+\lambda (w(j)-w(j-1))(l-j),
\]
 and get the desired result as follows
\[\begin{split}
&w(l)-w(j)+\lambda(j-l)f(j)\\
&\le
w(j)+\lambda (w(j)-w(j-1))(l-j)\\
&\qquad \quad~-w(j)+\lambda(j-l)f(j)\\
&=\lambda(l-j)(w(j)-w(j-1)-f(j))\le 0\,,
\end{split}
\]
where the last inequality holds because $f(j)\ge w(j)-w(j-1)$ (by assumption) and $l>j$.

Still within the region with $j+l>n$, and $l>j$, we conclude by considering the case of $j< n/2$. In this case we intend to show that any constraint identified by $(j,l)$ is implied by the corresponding constraint $(j,n-j)$. Since we have shown in part $\bf i)$ that the constraint $(j,n-j)$ is implied by the constraint $(j,j)$, this will conclude this part of the proof. We are therefore left to show that 
\[\begin{split}\\	
&w(n-j)+\lambda[jf(j)-(n-j)f(j+1)]\ge\\
&w(l)+\lambda[(n-l)f(j)-(n-j)f(j+1)],
\end{split}
\]
which is equivalent to 
\[
w(l)-w(n-j)+\lambda f(j)[(n-j)-l]\le 0.
\]
Since we are considering the case of $j< n/2$, it is $j< n-j$. Thus, by non-increasingness of $f$, it is $f(j)\ge f(n-j)$. Since $j+l>n$ implies $(n-j)-l<0$, we have 
\[
\begin{split}
&w(l)-w(n-j)+\lambda f(j)[(n-j)-l]\le \\
&w(l)-w(n-j)+\lambda f(n-j)[(n-j)-l]
\,.	
\end{split}
\]
Therefore the desired claim is shown if we can prove that $w(l)-w(n-j)+\lambda f(n-j)[(n-j)-l]\le 0$. Upon defining $q=n-j$, this reads as 
$w(l)-w(q)+\lambda (q-l)f(q)\le 0$. We observe that this claim is exactly that of \eqref{eq:provedfirst} (where $j$ has been substituted with $q$, and it is $l>q$), and has already been shown in the first part of {\bf ii)} (without ever using the fact that $j\ge n/2$). Therefore $w(l)-w(n-j)+\lambda f(n-j)[(n-j)-l]\le 0$ for all $j,l\in [0,n]$ with $j+l>n$, $l>j$, and $j< n/2$.
\end{enumerate}
The steps i) and ii) showed that $W^\star$ in \cite[Eq. \eqref{eq:formulacor1}]{part1Paccagnan2018} can be equivalently computed as
\[
\small
	\begin{split}
	W^\star &= \min_{\lambda\in\mb{R}_{\ge0},\,\mu\in\mb{R}}~ \mu  \\[0.1cm]
	&\,\text{s.t.}\hspace*{7mm}\lambda \ge 1
	\\
	&\!\qquad\mu w(j)\!\ge \!w(l)+\lambda[j f(j) \!-\!l f(j\!+\!1) ]\\%1\le j+l\le k\\
	&~\hspace*{25mm} \forall j,l\in[0, n]~{s.t.}~ j\ge l ~\text{and}~ 1\le j+l\le n,\\[0.15cm]
	&\!\qquad \mu w(j)\!\ge\! w(l)\!+\!\lambda[(n\!-\!l) f(j) \!-\!(n\!-\!j) f(j\!+\!1) ]\\
	&~\hspace*{25mm} \forall j,l\in[0, n]~{s.t.}~ j\ge l ~\text{and}~~~~~~ j+l> n.
	\end{split}
\]
Observe that every constraint appearing in the previous program is indexed by $(j,l)$, and $j=0$ is not included. All these constraints can be compactly written as $\mu\ge b_{jl}+ a_{jl}\lambda$, upon defining $b_{jl}\coloneqq w(l)/w(j)$  (by assumption $w(j)>0$, $j\in [n]$), and consequently 
\[
a_{jl}\coloneqq
\begin{cases}
[j f(j)\!-\!l f(j\!+\!1)]/w(j)\hspace*{16.5mm} 1\le j+l\le n,\\
[(n\!-\!l) f(j) \!-\!(n\!-\!j) f(j\!+\!1)]/w(j) \qquad j+l>n.
\end{cases}
\]
Therefore $W^\star$ can be computed as
\[
\small
	\begin{split}
	W^\star &= \min_{\lambda\ge 1,\,\mu\in\mb{R}}~ \mu  \\[0.1cm]
	&\,\text{s.t.}~ \,\mu\ge b_{jl}+ a_{jl}\lambda\qquad\forall j,l\in[0, n],~\text{s.t.}~j\ge l,~j\ge 1.
	\end{split}
\]
Observe that, when $j\ge 1$ and $j\ge l$, it holds $a_{jl} \ge 0$. 
Indeed since $f(j)$ is non-increasing, for $1\le j+l \le n$ one has $a_{jl}=[j f(j)\!-\!l f(j\!+\!1)]/w(j) \ge (j-l)f(j)/w(j)\ge 0$. Similarly for $j+l> n$. 
Thus the optimal choice is to select $\lambda$ as small as possible i.e. $\lambda^\star=1$. It follows that $W^\star$ becomes 
\[
\small
	\begin{split}
	W^\star &= \min_{\mu\in\mb{R}}~ \mu  \\[0.1cm]
	&\,\text{s.t.}~ \,\mu w(j)\!\ge \!w(l)+j f(j) \!-\!l f(j\!+\!1) \\%1\le j+l\le k\\
	&~\hspace*{25mm} \forall j,l\in[0, n]~{s.t.}~ j\ge l ~\text{and}~ 1\le j+l\le n,\\[0.15cm]
	&\!\qquad \mu w(j)\!\ge\! w(l)\!+\!(n\!-\!l) f(j) \!-\!(n\!-\!j) f(j\!+\!1) \\
	&~\hspace*{25mm} \forall j,l\in[0, n]~{s.t.}~ j\ge l ~\text{and}~~~~~~ j+l> n.
	\end{split}
\]
While the constraints with $j=0$ are not included in the above linear program, in the following we eliminate also the constraints with $l=0$, $j\in[n]$. Indeed the constraints with $l=0$, $j\in[n]$ are redundant as the corresponding constraints with $l=1$, $j\in[n]$ are always tighter. This follows from 
\[
\mu w(j)\ge w(1)+jf(j)-f(j+1)\ge jf(j),\quad \forall j\le n-1,
\] 
where the last inequality is equivalent to $f(j+1)\le w(1)=1$ which holds as $f(j)$ is assumed to be non-increasing and $f(1)=1$.
Similarly, for $j=n$ it is
\[
\mu w(n)\ge w(1)+(n-1)f(n)\ge  nf(n),
\]
where the last inequality is equivalent to $f(n)\le w(1)=1$, which can be proved as in the above. Rearranging the indices we get the desired result
\[
\small
	\begin{split}
	W^\star &= \min_{\mu\in\mb{R}}~ \mu  \\[0.1cm]
	&\,\text{s.t.}~ \,\mu w(j)\!\ge \!w(l)+j f(j) \!-\!l f(j\!+\!1) \\%1\le j+l\le k\\
	&~\hspace*{25mm} \forall j,l\in[n]~{s.t.}~ j\ge l ~~\text{and}~~  j+l\le n,\\[0.15cm]
	&\!\qquad \mu w(j)\!\ge\! w(l)\!+\!(n\!-\!l) f(j) \!-\!(n\!-\!j) f(j\!+\!1) \\
	&~\hspace*{25mm} \forall j,l\in[n]~{s.t.}~ j\ge l ~~\text{and}~~j+l> n.
	\end{split}
\]
\end{proof}
\section*{Proof of Corollary \ref{cor:SVandMC}}
\begin{proof}
	The proof is an application of Theorem \ref{thm:wconcavefwdecreas}.	%
	\setlist[enumerate]{leftmargin=*}
	\begin{enumerate}
	\item[\bf i)] Observe that $\fsv$ satisfies the assumptions of Theorem \ref{thm:wconcavefwdecreas} in that $\fsv(1)=1$ (because $w(1)=1$ by Assumption \ref{ass:fw=1}), $\fsv(j)={w(j)}/{j}$ is non-increasing (due to concavity of $w$), and it also holds that $\fsv(j)={w(j)}/{j}\ge w(j)-{w(j-1)}$ (trivially satisfied for $j=1$) while for $j> 1$ this is equivalent to  $w(j)/j\le w(j-1)/(j-1)$, which holds due to the concavity of $w$. Hence the result of Theorem \ref{thm:wconcavefwdecreas} applies and substituting $\fsv(j)=w(j)/j$ in \eqref{eq:poasubmodular} gives~$W^{\rm es}$.
	\item[\bf ii)]
	Observe that $\fmc$ satisfies the assumption of Theorem \ref{thm:wconcavefwdecreas} in that $\fmc(1)=1$ (because $w(1)=1$ by Assumption \ref{ass:fw=1}), $\fmc(j)=w(j)-w(j-1)$ is non-increasing (due to concavity of $w$).
	We conclude by proving that the constraints indexed with $l\le j\in[n]$ are not needed and it is enough to consider $j=l\in[n]$, so that $W^{\rm mc}$ is as given in Corollary \ref{cor:SVandMC}, {\bf ii)}. To do so, we show that for any $l<j$ the most binding constraint is given by $l=j$.
	
	In the case of $l<j$ and $j+l\ge n$ we intend to show
	\[\begin{split}
&w(j)+ (n-j)[\fmc(j)-\fmc(j+1)]\ge\\
& w(l)+(n-l)\fmc(j)-(n-j)\fmc(j+1),
\end{split}
\]
where the left hand side is obtained setting $l=j$. This can be equivalently written as
	\be
	\begin{split}
	&w(l)-w(j)+(j-l)\fmc(j)=\\
	&w(l)-w(j)+(j-l)(w(j)-w(j-1))
	\le 0\,,
	\end{split}
	\label{eq:proofformula2bis}
	\ee
	which holds since, by concavity of $w$ and $l<j$, it is $w(j)\ge w(l)+(j-l)(w(j)-w(j-1))$.

	In the case of $l<j$, $j+l< n$, and $j\le n/2$ we want to prove that
\[
\begin{split}
&w(j)+j[\fmc(j)-\fmc(j+1)]\ge\\
& w(l)+j\fmc(j)-l\fmc(j+1)\,,
\end{split}
\]
where the left hand side is obtained setting $l=j$.
The previous is equivalent to 
	\[
	w(l)-w(j)+(j-l)(w(j+1)-w(j))\le 0\,,
	\]
	which holds by concavity of $w$ and $l<j$ (similarly to how \eqref{eq:proofformula2bis} is shown).
	
	{For $l<j$, $j+l< n$, and $j> n/2$ we intend to prove that the constraints indexed by $(j,l)$ is implied by the corresponding constraint $(j,n-j)$. Since we have shown (in the first case of the proof) %
	that the constraint $(j,n-j)$ is implied by the constraint $(j,j)$, this will conclude the argument.
	We are thus left to show that 
	\[
	\begin{split}
	&w(n-j)+j\fmc(j)-(n-j)\fmc(j+1)\ge\\
	&w(l)+j\fmc(j)-l\fmc(j+1)\,,	
	\end{split}
	\]
	which can be equivalently written as
	\[
	w(l)-w(n-j)+[(n-j)-l]\fmc(j+1)\le 0\,.
	\]
	Since $j>n/2$, it is $j+1>n-j$, and by non-increasingess of $\fmc$ it holds $\fmc(j+1)\le \fmc(n-j)$. Therefore, since $n-j-l>0$, we have
	\[
	\begin{split}
	&w(l)-w(n-j)+[(n-j)-l]\fmc(j+1)\le \\
	&w(l)-w(n-j)+[(n-j)-l]\fmc(n-j)\,.
	\end{split}
	\]
	We are thus left to show that $w(l)-w(n-j)+[(n-j)-l]\fmc(n-j)\le 0$. Upon defining $q=n-j$, $q>l$ this inequality reads as $w(l)-w(q)+\fmc(q)(q-l)\le 0$, which holds by concavity of $w$ and $l<q$ (similarly to \eqref{eq:proofformula2bis}).
	}
	
		Hence, the price of anarchy of $\fmc$ is governed by $W^\star$ as in Theorem \ref{thm:wconcavefwdecreas}, where we set $f=\fmc$ and fix $j=l$. This gives the following expression
\[
	\small
	W^{\rm mc} = 1 + \max_{j\in [n]}\biggl\{\min(j,n-j)\left[\frac{\fmc(j)}{w(j)}-\frac{\fmc(j+1)}{w(j)}\right]\biggl
	\},
\]
	which reduces to the expression for $W^{\rm mc}$ in the claim, upon substituting $\fmc$ with its definition.
\end{enumerate}
\end{proof}
\section*{Proof of Theorem \ref{thm:poageneralcovering}}
\begin{proof} The proof is a specialization of the general result obtained in 
\cite[Thm. \ref{thm:dualpoa}]{part1Paccagnan2018} to the case of set covering problems.
We divide the study in three distinct cases, as in the following
\[ 
\cdue: \begin{cases}
a+x=0\\
	b+x\neq 0
\end{cases}
\quad
\ctre: \begin{cases}
a+x\neq0\\
	b+x= 0
\end{cases}
\quad
\]
\[
\cquattro: \begin{cases}
a+x\neq0\\
	b+x\neq 0
\end{cases}
\]
In case $\cdue$ it must be $a=x=0$, $b\neq 0$ and the constraints read as 
\[
\lambda\ge \frac{1}{b}\,.
\]
The most binding one is obtained for $b=1$, i.e. it suffices to have $\lambda\ge 1$ in order to guarantee $\lambda\ge 1/b$.  %
In case $\ctre$ it must be $b=x=0$, $a\neq 0$. The constraints read as 
\[
\mu\ge \lambda af(a)\quad\forall~a\in[n].\]
In case $\cquattro$, since $a+x\neq 0 $ and $b+x\neq 0$, the constraints become 
\[
\mu\ge 1+\lambda[af(a+x)-bf(a+x+1)]\,.
\]

If $x=0$, then $a,\,b>0$ and the previous inequality reads
\[
\mu\ge 1+\lambda[af(a)-bf(a+1)]\quad a+b\in[n],.
\]
The most constraining inequality is obtained for $b$ taking the smallest possible value, that is $b=1$. Thus $0< a\le n-1$. Consequently when $x=0$, it suffices to have 
\[
\mu\ge 1+\lambda[af(a)-f(a+1)]\quad\forall a\in[n-1]\,.
\]

If $x\neq0$, the most binding constraint is obtained for $b=0$. In such case, $0< a+x\le n$  and the constraints read as 
\[
\mu\ge 1+\lambda af(a+x)\quad\forall a\in[n]\,.
\]
For ease of readability, we introduce the variable $j\coloneqq a+x$ and use $j$ and $x$ instead of $a$ and $x$. With this new system of indices the feasible region becomes $0< j \le n$ and $j-x\ge0$, $x> 0$. The constraints read as
\[
\mu\ge 1+\lambda (j-x)f(j)%
\]
and the most binding is trivially obtained for $x=1$, reducing the previous to
\[
\mu\ge 1+\lambda (j-1)f(j)\quad\forall ~j\in[n]\,.
\]

This guarantees that the program in \cite[Eq. \eqref{eq:generalbound}]{part1Paccagnan2018} is equivalent to
\[
\begin{split}
W^\star=&\min_{\lambda\in\mb{R}_{\ge0},\,\mu\in\mb{R}}~ \mu \\
&\,\text{s.t.}\quad \lambda \ge {1}\\
& \qquad~ \mu \ge \lambda jf(j)\quad j\in[n]\\
&\qquad~\mu\ge1+ \lambda(jf(j)-f(j+1))\quad j\in[n-1]\\
&~\qquad \mu\ge1+ \lambda (j-1)f(j)\quad j\in[n]\,.
\end{split}
\]
Amongst the last three set of constraints, the tightest constraint always features a positive coefficient multiplying $\lambda$. Indeed the only term multiplying $\lambda$ that could take negative values is $jf(j)-f(j+1)$, but every time this is negative, the constraints $\mu\ge1+ \lambda (j-1)f(j)$ are tighter. %
It follows that the solution consists in picking $\lambda$ as small as possible, that is in choosing $\lambda^\star = 1$. The program becomes 
\[
\begin{split}
W^\star=&\min_{\mu\in\mb{R}}~ \mu \\
&\,\text{s.t.}\quad \mu \ge  jf(j)\quad j\in[n]\\
&\qquad~\mu\ge1+ jf(j)-f(j+1) \quad j\in[n-1]\\
&~\qquad \mu\ge1+ (j-1)f(j)\quad j\in[n]\,.
\end{split}
\]
We conclude with a little of cosmetics: the first and third set of inequalities run over $j\in[n]$, while the second one has $j\in[n-1]$. Observe that the first and the third condition evaluated at $j=1$ read both as $\mu\ge 1$. This condition is implied by the last set of condition with $j=2$, indeed it reads as $\mu\ge 1+f(2)\ge1$ since we assumed $f$ non-negative. Thus the first and third conditions can be reduced to $j\in[2, n]$. Shifting the indices down by one, we get 
\[
\begin{split}
W^\star=&\min_{\mu\in\mb{R}}~ \mu \\
&\,\text{s.t.}\quad \mu \ge  (j+1)f(j+1)\quad j\in[n-1]\\
&\qquad~\mu\ge1+ jf(j)-f(j+1) \quad j\in[n-1]\\
&~\qquad \mu\ge1+ jf(j+1)\quad j\in[n-1]\,,
\end{split}
\]
from which we get the analytic expression in \eqref{eq:Wstarsetcoveringg}, i.e.,
\[
W^\star = 1+\!\!\max_{j\in [n-1]}\!\{(j+1)f(j+1)-1, jf(j)-f(j+1),jf(j+1)\}\,.
\]
\end{proof}
\section*{Proof of Corollary \ref{cor:backtogair}}
\begin{proof}
Thanks to Theorem \ref{thm:poageneralcovering}, the value $W^\star$ and consequently the price of anarchy can be computed as
\[\small
W^\star \!= \!\!\max_{j\in [n-1]}\!\{(j+1)f(j+1),1+ jf(j)-f(j+1),1+jf(j+1)\}.\]
We will show that when $f$ is non-increasing, fewer constraints are required, producing exactly \eqref{eq:Wstarcoveringgair}.

First observe that $f$ being non-increasing implies $(j+1)f(j+1)= f(j+1)+j f(j+1)\le f(1)+jf(j+1)= 1+ j f(j+1)$, so that the first set of conditions %
is implied by the third. Hence
\[
W^\star = 1+\max_{j\in [n-1]}\{jf(j)-f(j+1),\,jf(j+1)\}\,.\]
We now verify that the first set of remaining conditions implies all the conditions in the second set, but not the last one:
\[
\mu\ge 1+jf(j)-f(j+1)\ge 1+jf(j)-f(j)=1+(j-1)f(j)\,,
\]
$\forall j\in[n-1]$ that is, all conditions $\mu \ge 1+ j f(j+1)$ are satisfied for $j\in [n-2]$. 
Thus, it suffices to require $\mu-1\ge  jf(j)-f(j+1)$ and $\mu-1\ge (n-1)f(n) $ for all $j\in [n]$ and the result in \eqref{eq:Wstarcoveringgair} follows.
\end{proof}

\section*{Proof of Theorem \ref{thm:convex}}
\begin{proof}
The proof is a specialization of the general result obtain in Theorem \ref{thm:dualpoa}. We divide the study in the same three cases used for the proof of Theorem \ref{thm:poageneralcovering}. 

\noindent In case $\cdue$, the constraints read as 
\[
w(b)-\lambda b\le 0 \iff \lambda \ge \frac{w(b)}{b},
\] 
the most constraining of which is given for $b=n$ as $w(b)$ is convex. Thus it must be 
\[
\lambda\ge \frac{w(n)}{n}\
\]

\noindent In case $\ctre$, the constraints read as
\[
\lambda a f(a)\le \mu w(a) \iff \mu\ge \lambda a \frac{f(a)}{w(a)}.
\]
In case $\cquattro$, the constraints read as
\[
\mu\ge \frac{w(b+x)}{w(a+x)}+\lambda \biggl[a \frac{f(a+x)}{w(a+x)}-b\frac{f(a+x+1)}{w(a+x)} \biggr ]\,.
\]
In order to conclude, we will show that the constraints obtained from $\cdue$ and $\ctre$ imply all the conditions stemming from $\cquattro$.
To do so observe that 
\[
\begin{split}
  &\frac{w(b+x)}{w(a+x)}+\lambda \biggl[a\frac{f(a+x)}{w(a+x)}-b\frac{f(a+x+1)}{w(a+x)}\biggr ] \\
= &\frac{1}{w(a+x)}\biggl[w(b+x)-\lambda bf(a+x+1) +\lambda af(a+x)\biggr ]\\
\le &\frac{1}{w(a+x)}\biggl[\lambda (b+x)-\lambda b +\lambda af(a+x)\biggr ]\\
 = &\frac{1}{w(a+x)}\left[ x\lambda +\lambda af(a+x)\right]\le \lambda (a+x)\frac{f(a+x)}{w(a+x)}
\end{split}
\]
From first to second line is rearrangement. From second to third is due to $f(a+x+1)\ge 1$ and to $w(b+x)\le \frac{w(n)}{n}(b+x)\le \lambda (b+x)$  where the first inequality holds because of convexity of $w$ and the second inequality follows from $\cdue$ (i.e., from $\lambda \ge \frac{w(n)}{n}$). The last inequality follows from $1 \le f(a+x)$.

 The previous series of inequalities have demonstrated that if $\mu\ge \lambda a f(a)/w(a)$ as required by condition $\ctre$, and if $\lambda\ge  \frac{w(n)}{n}$ as required by condition $\cdue$, then $\mu \ge \lambda (a+x)\frac{f(a+x)}{w(a+x)}\ge\frac{w(b+x)}{w(a+x)}+\lambda \left[a \frac{f(a+x)}{w(a+x)}-b\frac{f(a+x+1)}{w(a+x)} \right]$ i.e. conditions $\cquattro$ are all satisfied.
 
It follows that $W^\star$ and consequently the price of anarchy is easily obtained as 
\be
\begin{split}
&W^\star = \min_{\lambda\in\mb{R}_{\ge0},\,\mu\in\mb{R}}~ \mu\\
&\text{s.t.~}\mu\ge  \lambda j \frac{f(j)}{w(j)}\quad\forall j \in[n]\\
&~~~~\,\lambda\ge \frac{w(n)}{n}\,.
\end{split}
\ee
The solution is $\lambda^\star = \frac{w(n)}{n}$, $\mu^\star = \lambda^\star \max_{j\in [n]}j f(j)/w(j)$, which gives a price of anarchy of 
\[
\poa{(f,w,n)}= \frac{n}{w(n)} \frac{1}{\max_{j\in[n]} j \cdot \frac{f(j)}{w(j)}}\,.
\]
The optimality of $\fsv$ follows from the fact that $\max_{j\in[n]} j \cdot \fsv(j)/w(j)=1$ is the smallest achievable value.
\end{proof}
\bibliographystyle{IEEEtran}
\bibliography{biblio_poaLP}

\vspace*{-5mm}
\begin{IEEEbiography}
[{\includegraphics[width=1in,height=1.25in,clip,keepaspectratio]{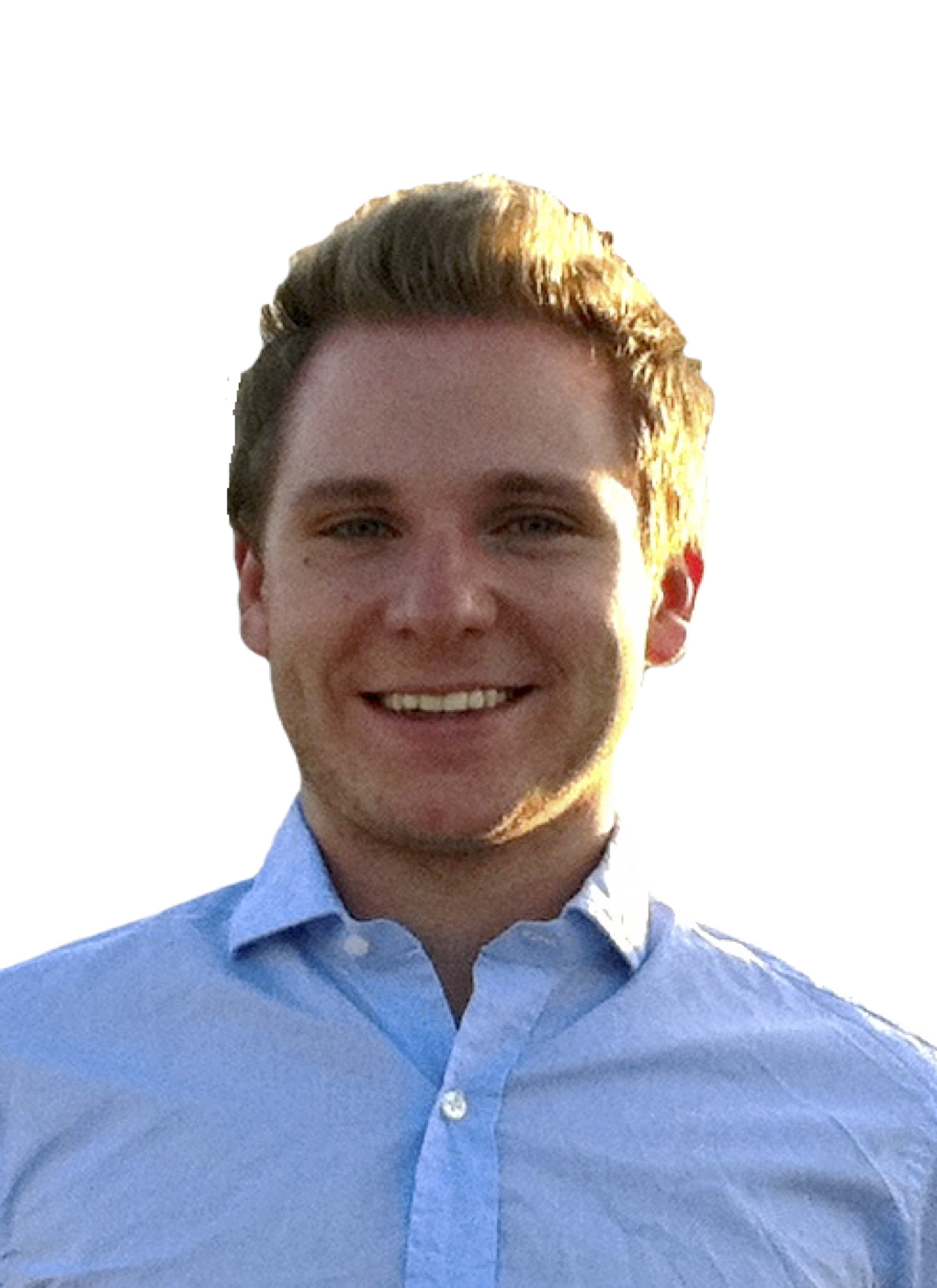}}]
{Dario Paccagnan} received the Bs.C. (Hons.) and M.Sc. (Hons.) degrees in aerospace engineering from the University of Padova, Padova, Italy, in 2011 and 2014, respectively. He also received the M.Sc. (Hons.) degree in in mathematical modeling and computation from the Technical University of Denmark, Kongens Lyngby, Denmark, in 2014, and the Ph.D. degree in optimization and control from ETH Zurich, Zurich, Switzerland, in 2018.

He is an Assistant Professor (U.K. Lecturer) with the Department of Computing, Imperial College London, London, U.K., since the Fall 2020. Before that, he was a Postdoctoral Fellow with the Center for Control, Dynamical Systems and Computation, University of California, Santa Barbara, Santa Barbara, CA, USA. 
His research interests are at the interface between game theory and control theory, with a focus on the design of behavior-influencing mechanisms for socio-technical systems.

Dr. Paccagnan was recognized with the ETH Medal for his doctoral work, the SNSF Early Postdoc Mobility Fellowship, and the SNSF Doc Mobility Fellowship. He was also a finalist for the 2019 EECI Best Ph.D. Thesis Award.
\end{IEEEbiography}
\vspace*{-7mm}
\begin{IEEEbiography}
[{\includegraphics[width=1in,height=1.25in,clip,keepaspectratio]{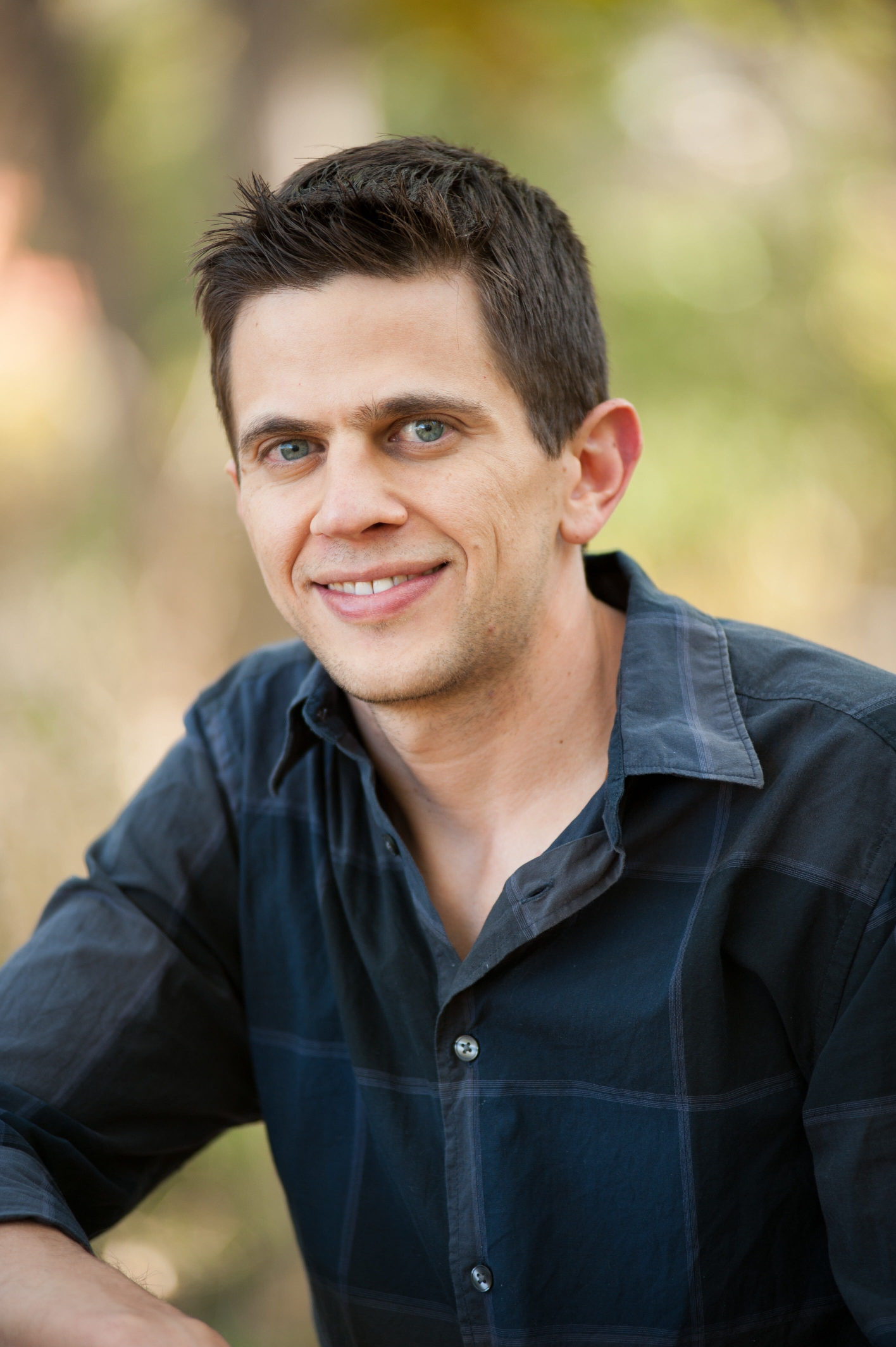}}]
{Jason Marden} is an Associate Professor in the Department of Electrical and Computer Engineering at the University of California, Santa Barbara. Jason received a BS in Mechanical Engineering in 2001 from UCLA, and a PhD in Mechanical Engineering in 2007, also from UCLA, under the supervision of Jeff S. Shamma, where he was awarded the Outstanding Graduating PhD Student in Mechanical Engineering. After graduating from UCLA, he served as a junior fellow in the Social and Information Sciences Laboratory at the California Institute of Technology until 2010 when he joined the University of Colorado. Jason is a recipient of the NSF Career Award (2014), the ONR Young Investigator Award (2015), the AFOSR Young Investigator Award (2012), the American Automatic Control Council Donald P. Eckman Award (2012), and the SIAG/CST Best SICON Paper Prize (2015). Jason's research interests focus on game theoretic methods for the control of distributed multiagent systems.
\end{IEEEbiography}
\vfill
\end{document}